\documentclass[10pt,twocolumn,aps,prd,nofootinbib,superscriptaddress]{revtex4-1}

\usepackage[T1]{fontenc}
\usepackage[utf8]{inputenc}
\usepackage{amssymb}
\usepackage{mathtools}
\usepackage{fontawesome}
\usepackage[colorlinks,pdfusetitle]{hyperref}
\hypersetup{allcolors=[rgb]{1,0.56,0}}
\usepackage{breakurl} 

\newcommand{\orcid}[1]{\href{https://orcid.org/#1}{#1}}

\newcommand{\e}[1]{\times10^{#1}}

\begin{document}

\title{Visible Decay of Astrophysical Neutrinos at IceCube}

\author{Asli Abdullahi}
\email{asli.abdullahi@durham.ac.uk}
\thanks{\orcid{0000-0002-6122-4986}}
\affiliation{Institute for Particle Physics Phenomenology, Department of Physics, Durham University, Durham DH1 3LE, U.K.}

\author{Peter B.~Denton}
\email{pdenton@bnl.gov}
\thanks{\orcid{0000-0002-5209-872X}}
\affiliation{High Energy Theory Group, Physics Department, Brookhaven National Laboratory, Upton, NY 11973, USA}

\date{May 14, 2020}

\begin{abstract}
{\centering\href{https://github.com/PeterDenton/Astro-Nu-Decay}{\large\faGithub}\\}
Neutrino decay modifies neutrino propagation in a unique way; not only is there flavor changing as there is in neutrino oscillations, there is also energy transport from initial to final neutrinos.
The most sensitive direct probe of neutrino decay is currently IceCube which can measure the energy and flavor of neutrinos traveling over extragalactic distances.
For the first time we calculate the flavor transition probability for the cases of visible and invisible neutrino decay, including the effects of the expansion of the universe, and consider the implications for IceCube.
As an example, we demonstrate how neutrino decay addresses a tension in the IceCube data.
We also provide a publicly available code to calculate the effect of visible decay.
\end{abstract}

\maketitle

\section{Introduction}
IceCube's discovery of a new flux of astrophysical neutrinos at TeV-PeV energies \cite{Aartsen:2013bka,Aartsen:2013jdh,Aartsen:2016xlq} of extragalactic origin \cite{Denton:2017csz,Aartsen:2017ujz} opens up the window to many new physics probes.
Due to the extremely long distances involved in neutrino propagation, this flux of neutrinos provides a unique opportunity to probe new physics models, in particular neutrino decay.
While neutrinos do decay in the Standard Model (SM)\footnote{While the SM does not tell us how neutrino masses are generated, once they are neutrinos will decay. The details of neutrino mass generation do not affect this work.}, the decay is highly suppressed and unapproachable by any foreseen experiment \cite{Petcov:1976ff,Marciano:1977wx}.
In this work, we consider the well-studied Majoron model, which postulates the existence of a singlet Higgs-like scalar with non-zero lepton number \cite{Chikashige:1980ui,Gelmini:1980re,Schechter:1981cv,Acker:1991ej}.
Such a scalar could generate the Majorana mass of a right-handed sterile neutrino through a lepton number violating vacuum expectation value.
The spontaneous breaking of the global lepton number symmetry would then provide us with a massless Goldstone boson, $\phi$ - the \textit{Majoron}.
Through mixing between active states and the sterile, we could obtain off-diagonal couplings between the neutrino mass eigenstates and $\phi$. Generically, we are interested in Lagrangian terms of the form
\begin{equation}
\mathcal{L} \supset \frac{g_{ij}}2\bar\nu_j \nu_i \phi+\frac{g'_{ij}}2\bar\nu_ji\gamma_5\nu_i\phi+{\rm h.c.}\,,
\label{eq:lagrangian}
\end{equation}
where the $g^{(\prime)}_{ij}$ are the scalar (pseudo-scalar) couplings.
While Majorons are potential dark matter candidates \cite{Rothstein:1992rh,Berezinsky:1993fm,Frigerio:2011in,Queiroz:2014yna,Heeck:2017xbu,Heeck:2017kxw,Garcia-Cely:2017oco,Brune:2018sab,Heeck:2018lkj,Heeck:2019guh,Abe:2020dut}, in those scenarios the Majorons are too heavy for neutrinos to decay.

Such a term in the Lagrangian would then induce the decays of the light neutrino mass eigenstates, $\nu_i \to\nu_j + \phi$.
Both $\nu_i\to\nu_j$ and $\nu_i\to\bar\nu_j$ decays can occur.
In the case of Dirac neutrinos, the $\nu_i\to\bar\nu_j$ decays result in neutrinos in the mostly sterile direction (assuming a light sterile) which are undetectable (unless there is a secondary $\bar\nu_j\to\nu_k$ decay), while in the case of Majorana neutrinos such a channel is detectable leading to the possibility of differentiating Majorana neutrinos from Dirac neutrinos by the measurement of neutrino decay \cite{Coloma:2017zpg,deGouvea:2019goq,Funcke:2019grs}.
In this work we will focus on the Majorana case, although our results can be easily extended to the Dirac case as well.
Phenomenologically there are two kinds of decay, each of which leave distinct imprints on the measured flux.

\textbf{Invisible Decay}
In these decays, the final state neutrino is not observable in the detector, either because it is sterile or because its energy is too low to produce a signal through scattering.
Invisible decay results in a depletion of the number of events below a fiducial energy given by the coupling and distance traveled.
Typically this results in fewer $\nu_\mu$'s and $\nu_\tau$'s assuming the normal mass ordering.
While we include results for invisible neutrino decay for comparison with other works in the literature, we are agnostic about the model details and focus mostly on the effect of visible decay.

\textbf{Visible Decay}
Here, the final state neutrino is observable in the detector.
The decay of neutrinos during propagation appears as a depletion of the heavier mass eigenstates as with invisible decay.
Simultaneously, we get an increase in the number of lightest eigenstates, and hence more $\nu_e$'s in the normal mass ordering.
Another signature of visible decay is in the energy of the observed neutrinos.
As the neutrinos produced in decay have less energy than the source neutrinos, the increase in lighter mass eigenstates is in lower energy bins than the parent neutrinos.

Neutrino decay is constrained in a wide range of experiments.
In general, the longer the baseline and the lower the neutrino energy, the stronger the constraint.
The weakest constraints come from atmospheric, long-baseline accelerator, and reactor neutrinos which are mainly sensitive to $\nu_3$ decays \cite{Barger:1998xk,Fogli:1999qt,Meloni:2006gv,Maltoni:2008jr,GonzalezGarcia:2008ru,Abrahao:2015rba,Pagliaroli:2016zab,Choubey:2017eyg,Coloma:2017zpg,Choubey:2017dyu,Choubey:2018cfz,deSalas:2018kri,Ascencio-Sosa:2018lbk,Tang:2018rer,Porto-Silva:2020gma,Ghoshal:2020hyo} or in conjunction with sterile neutrinos \cite{Moss:2017pur}.
Hints of $\nu_3$ decay has been found in NOvA and T2K data \cite{Gomes:2014yua,Choubey:2018cfz}.
The next strongest constraints come from solar neutrinos and apply mostly to $\nu_2$ \cite{Bahcall:1972my,Raghavan:1987uh,Berezhiani:1991vk,Joshipura:1992vn,Acker:1992eh,Berezhiani:1992ry,Berezhiani:1993iy,Choubey:2000an,Bandyopadhyay:2001ct,Beacom:2002cb,Joshipura:2002fb,Bandyopadhyay:2002qg,Berryman:2014qha,Picoreti:2015ika,Aharmim:2018fme,Funcke:2019grs}.
Next, neutrino decay has been considered at IceCube using the astrophysical neutrino flux which constrains the cases of all states decaying simultaneously \cite{Beacom:2002vi,Beacom:2003nh,Baerwald:2012kc,Dorame:2013lka,Pagliaroli:2015rca,Bustamante:2015waa,Bustamante:2016ciw} or only the heavy two states \cite{Denton:2018aml,Bustamante:2020niz}.
In these cases, however, a full treatment of visible decay including cosmology and energy transfer has not been considered.

Interestingly, ref.~\cite{Denton:2018aml} found evidence at $>3$ $\sigma$ that invisible partial neutrino decay is preferred over the SM due to a tension in the data related to the differences in the track and cascade spectra\footnote{The track to cascade spectral difference was also investigated in \cite{Palladino:2019pid} which pointed out that a neutron decay source scenario somewhat relaxes the tension, although such astrophysical models are quite unlikely \cite{Anchordoqui:2014pca}.}.
The strongest direct constraints on invisible neutrino decay come from SN1987A which constrains $\bar\nu_e$ decay \cite{Hirata:1987hu,Frieman:1987as,Berezhiani:1989za,Kachelriess:2000qc}.

\begin{figure}
\centering
\includegraphics[width=\columnwidth]{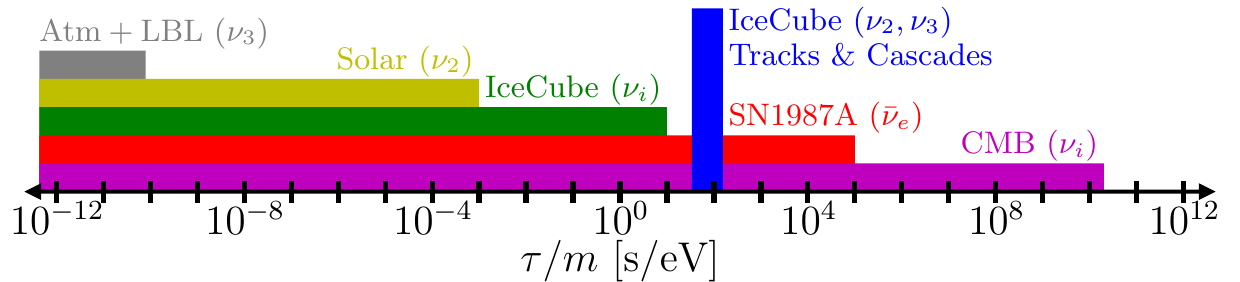}
\caption{The constraints on invisible neutrino decay from various sources as a function of the lifetime over the mass ($\tau/m$) \cite{Abrahao:2015rba,Aharmim:2018fme,Bustamante:2016ciw,Hirata:1987hu,Escudero:2019gfk}.
Each constraint only applies to certain mass/flavor eigenstates except for the CMB constraint which is essentially independent of which or how many states decay, and the IceCube constraint which only applies when all states decay with the same lifetime.
The vertical blue line is a hint of invisible partial neutrino decay from IceCube data \cite{Denton:2018aml}.}
\label{fig:schematic}
\end{figure}

While the above constraints are directly related to the depletion of neutrinos, it is also possible to constrain the process of neutrino decay via the $\nu\nu\to\phi\phi$ diagram in the early universe leading to very tight bounds \cite{Hannestad:2004qu,Hannestad:2005ex,Bell:2005dr,Basboll:2008fx,Archidiacono:2013dua,Escudero:2019gfk}.
Big bang nucleosynthesis constraints do not depend on neutrinos explicitly decaying while the cosmic microwave background constraints were derived assuming neutrinos decay invisibly.
It is also possible, in principle, to construct a model that evades the early universe constraints and still predicts late time neutrino decay, for one such realization see ref.~\cite{Dvali:2016uhn}.
In addition, typical models that predict neutrino decay (such as in eq.~\ref{eq:lagrangian}) can also be probed in other environments such as supernova, neutrinoless double beta decay, and meson decay \cite{Dror:2020fbh}.
These constraints are summarized in fig.~\ref{fig:schematic}.

In addition, neutrino decay can be further probed in the future in the case of a galactic supernova \cite{Tomas:2001dh,Lindner:2001th,Ando:2004qe,deGouvea:2019goq}, a measurement of the diffuse supernova neutrino background \cite{Ando:2003ie,Beacom:2004yd,Fogli:2004gy}, improved solar neutrino measurements \cite{Huang:2018nxj}, or a measurement of the cosmic neutrino background \cite{Long:2014zva};
see ref.~\cite{Arguelles:2019xgp} for a review of new physics searches in upcoming neutrino experiments.

Due to the large mixing in the neutrino sector, neutrino decay may not significantly deplete the flux of neutrinos from SN1987A \cite{Frieman:1987as,Denton:2018aml} making IceCube the most sensitive direct probe of neutrino decay.
In addition, since IceCube is sensitive to high energy neutrinos well above the $\tau$ production threshold, IceCube is sensitive to six of the nine flavor changing channels $P_{ee},P_{\mu e},P_{e\mu},P_{\mu\mu},P_{e\tau},P_{\mu\tau}$ (it is unlikely that $\nu_\tau$'s are produced in high energy astrophysical sources) unlike supernova neutrino probes which can only detect $\nu_e$ and thus are sensitive to only three flavor changing channels: $P_{ee},P_{\mu e},P_{\tau e}$.
The sensitivity to different flavors provides a handle to break degeneracies with astrophysical uncertainties.
In this article we further explore the phenomenology of neutrino decay including the regeneration component relevant for visible neutrino decay.

This paper is set out as follows:
in section \ref{sec:formalism}, we describe in more detail the different decay scenarios, focusing on visible decay, and set up the formalism required to study the impact of neutrino decay on the flavor ratios.
We also consider the added effect of cosmology on the observed spectrum of the neutrinos.
We then analytically evaluate the integrals in some cases and take some interesting limits in section \ref{sec:analytic}.
In section \ref{sec:results}, we present the results of implementing the formalism in section \ref{sec:formalism}.
We study how the parameters in our model, namely our couplings with the Majoron, the absolute mass scale of the neutrinos, and the energy spectrum at the source, impacts visibility at Earth.
We also suggest some benchmark points that IceCube, and future experiments, could probe.
We discuss the possibility of mitigating the $>3$ $\sigma$ tension in the track and cascade events observed at IceCube in section \ref{sec:icecube}, and finish up with a discussion of our results and our conclusions in sections \ref{sec:discussion} and \ref{sec:conclusions}.
The code for this article is publicly available and can be found at \href{https://github.com/PeterDenton/Astro-Nu-Decay}{\texttt{github.com/PeterDenton/Astro-Nu-Decay}} \cite{peter_b_denton_2020_3826887} and the data files for all the figures can be found at \burlalt{peterdenton.github.io/Data/Visible\_Decay/index.html}{https://peterdenton.github.io/Data/Visible_Decay/index.html}.

\section{Neutrino Decay Formalism}\label{sec:formalism}
In this section we calculate the components of neutrino decay relevant for astrophysical neutrino experiments such as IceCube, KM3NeT, and Baikal \cite{Aartsen:2014njl,Adrian-Martinez:2016fdl,Avrorin:2018ijk}.
We assume the normal mass ordering scheme, $m_3 > m_2 > m_1$ due to the $\gtrsim3$ $\sigma$ preference from global fits \cite{Capozzi:2020qhw,Esteban:2018azc,Gariazzo:2018pei} where the mass eigenstates are defined by $|U_{e1}|^2>|U_{e2}|^2>|U_{e3}|^2$, see e.g.~\cite{Denton:2020exu}.
We take $\nu_1$ to be stable and let one or both of $\nu_3$ and $\nu_2$ decay, with the following possible decay channels: $\nu_3 \to \nu_2$, $\nu_3\to\nu_1$, $\nu_3\to\nu_2\to\nu_1$, and $\nu_2\to\nu_1$.
Our results can be extended to the inverted ordering in a straightforward fashion.
Also, since high energy neutrino experiments have very limited sensitivity to $\nu/\bar\nu$ discrimination\footnote{A few exceptions exist such as the Glashow resonance \cite{Glashow:1960zz,Anchordoqui:2004eb}, the inelasticity distribution \cite{Aartsen:2018vez}, and absorption, but the sensitivity of each is fairly limited.}, we consider both helicity conserving ($\nu\to\nu$ or $\bar\nu\to\bar\nu$) decays and helicity flipping ($\nu\to\bar\nu$ or $\bar\nu\to\nu$) decays.

\subsection{Invisible Decay}\label{ssec:invisible}
Invisible decay refers to the absence of decay products in the detector, either because they are sterile or too low energy to be detected (at lower energies the cross section is lower and, in the case of the high energy astrophysical flux, both the astrophysical flux and the atmospheric backgrounds are higher at lower energies).
This means that if we start with a flux of $\nu_\alpha$, due to the neutrinos decaying along the baseline, the flux of $\nu_\beta$ is expected to be less than that produced in oscillations alone. 

To begin, we define the total transition probability as the ratio of the observed spectrum at the Earth for a given channel, $\Phi_{\alpha\beta}^E$, over the initial spectrum at the source, $\Phi_{\alpha}^S$.
That is, the probability is the observed spectrum at the Earth divided by the spectrum at the source in the case of no oscillations or decay,
\begin{equation}\label{eq:prob_defn}
\bar P_{\alpha\beta}(E_f)\equiv\frac{\Phi_{\alpha\beta}^E(E_f)}{\Phi_{\alpha}^S(E_i)}\,,
\end{equation}
where we note that the initial energy $E_i$ is the same as the final energy for invisible decay.
Alternatively, the spectrum at the Earth can be simply computed as the probability calculated here multiplied by the initial spectrum.
We take a single power law spectrum (SPL) in the remaining work, 
\begin{equation}
    \Phi_{\alpha}^S(E_f)=\Phi^S_0E_f^{-\gamma}\,,
\end{equation}
where $\Phi^S_0$ is the flux normalization.
While a broken power law may also be considered \cite{Anchordoqui:2016ewn,Denton:2018aml,Aartsen:2020aqd}, it was not particularly preferred by IceCube data to an SPL source.
We note that the quantity $\bar P_{\alpha\beta}$ defined in eq.~\ref{eq:prob_defn} is not a true probability in that it can be larger than 1 depending on the spectrum. A clear definition of what we take to be the transition probability is necessary when the flux has non-trivial dependencies, as seen in sec.~\ref{subsec:cosmo} where the flux depends on the source redshift.

In the following, the energy $E$ refers to the energy at the source and that detected at Earth, as there is no change in energy in the SM, or in invisible decay. 

Starting with the SM transition probability for $\nu_\alpha\to\nu_\beta$, under the assumption of relativistic neutrinos with equal momentum,
\begin{multline}
    P_{\alpha\beta}^{\text{SM}}\left(E,L\right) = \left|U_{\alpha1}^*U_{\beta_1}+ U_{\alpha2}^*U_{\beta_2}e^{-i\frac{\Delta m^2_{21}}{2E}L}\right.\\
    +\left. U_{\alpha3}^*U_{\beta_3}e^{-i\frac{\Delta m^2_{31}}{2E}L}\right|^2\,.
\end{multline}
We can account for the effect of decay by making the substitution $\frac{\Delta m^2_{i1}}{2E} \to \frac{\Delta m^2_{i1}}{2E} -i\frac{\Gamma_{i}}{2}$ for unstable $\nu_i$, to obtain 
\begin{multline}\label{eq.inv}
    P_{\alpha\beta}^{\text{inv}}\left(E,L\right) = \left|U_{\alpha1}^*U_{\beta_1} + U_{\alpha2}^*U_{\beta_2}e^{-i\frac{\Delta m^2_{21}}{2E}L}e^{-\frac12\Gamma_2L}\right.\\
    +\left. U_{\alpha3}^*U_{\beta_3}e^{-i\frac{\Delta m^2_{31}}{2E}L}e^{-\frac12\Gamma_3L}\right|^2\,,
\end{multline}
where $\Gamma_i$ is the decay width for mass eigenstate $i$ in the lab frame.
The partial width in the lab frame for mass eigenstate $\nu_i$ to decay to $\nu_j$ of either helicity is,
\begin{multline}
\Gamma_{ij}=\frac{m_im_j}{16\pi E_i}\\
\times\left\{g_{ij}^2\left[f(x_{ij})+k(x_{ij})\right]+g^{\prime2}_{ij}\left[h(x_{ij})+k(x_{ij})\right]\right\}\,,
\label{eq:Gamma}
\end{multline}
where $E_i$ is the energy of the initial neutrino, $\nu_i$, $x_{ij}\equiv m_i/m_j$, and we have explicitly included both decay channels: $\nu_i\to\nu_j$ and $\nu_i\to\bar\nu_j$, and,
\begin{align}
f(x)&=\frac x2+2+\frac2x\log x-\frac2{x^2}-\frac1{2x^3}\,,\\
h(x)&=\frac x2-2+\frac2x\log x+\frac2{x^2}-\frac1{2x^3}\,,\\
k(x)&=\frac x2-\frac2x\log x-\frac1{2x^3}\,,
\end{align}
where the $f$ and $h$ equations represent the $\nu_i\to\nu_j$ case, for scalar and pseudo-scalar mediators respectively, and the $k$ equation represents the $\nu_i\to\bar\nu_j$ case, for both mediators \cite{Kim:1990km}.
The full width\footnote{The lifetime of neutrino $\nu_i$ its rest frame is $\tau_i=\frac{m_i}{E_i\Gamma_i}$ where we recall that $\Gamma_i$ is in the lab frame.} is then $\Gamma_i=\sum_j\Gamma_{ij}$.
Unless otherwise specified, we will consider the case of $g_{ij}=g'_{ij}$, the impact of which is shown in the middle left panel of fig.~\ref{fig:visible}.

As the baselines under consideration are very large, we can average over the oscillations to get,
\begin{equation}
\bar P_{\alpha\beta}^{\rm inv}=\bar P_{\alpha\beta}^{\rm SM}+\bar P_{\alpha\beta}^{\rm dep}\,,
\end{equation}
where the bar denotes the oscillation averaging since the states have decohered.
The SM component is,
\begin{equation}
\bar{P}_{\alpha\beta}^{\text{SM}} = \sum_i |U_{\alpha i}|^2 |U_{\beta i}|^2\,.
\end{equation}
and the depletion component is,
\begin{multline}
\bar{P}_{\alpha\beta}^{\text{dep}}\left(E,L\right)= - |U_{\alpha 2}|^2 |U_{\beta 2}|^2 (1-e^{-\Gamma_2 L})\\
- |U_{\alpha 3}|^2 |U_{\beta 3}|^2 (1-e^{-\Gamma_3 L})\,.
\end{multline}
We note that while $\bar P^{\rm dep}_{\alpha\beta}<0$, $\bar P^{\rm inv}_{\alpha\beta}>0$.
The smallest that $\bar P^{\rm dep}_{\alpha\beta}$ can be is $- |U_{\alpha 2}|^2 |U_{\beta 2}|^2- |U_{\alpha 3}|^2 |U_{\beta 3}|^2$, thus the smallest that $\bar P^{\rm inv}_{\alpha\beta}$ can be as a function of $g_{ij}$, $m_1$, and $L$ is $|U_{\alpha1}|^2|U_{\beta1}|^2$ since $\nu_1$ is stable. For the non-oscillation averaged depletion component, see appendix \ref{App. ssec. dep}.

\subsubsection*{Limiting Cases}
As a benchmark, we will assume that the neutrinos detected by IceCube result from full pion decay wherein the relative flux normalized to the $\nu_e$ flux is $(\Phi_{\nu_e}:\Phi_{\nu_\mu}:\Phi_{\nu_\tau})=(1:2:0)$, with all three neutrinos from the $\pi^\pm$ decay carrying nearly the same energy.
In the SM this results\footnote{Here and throughout we will use the global fit numbers from nu-fit 4.1 \cite{Esteban:2018azc}.} in a flavor ratio of $(1:1.152:1.117)$ after propagation.
It is interesting to consider some limits to demonstrate the effect of the decays on the flux observed at Earth. In the case where $\Gamma_2=0$ (that is, $g_{21}=0$) and $\Gamma_3\to\infty$, so that all $\nu_3$ decay at the source, we have 
$\bar{P}_{\alpha\beta}^{\text{inv}} \to |U_{\alpha 1}|^2 |U_{\beta 1}|^2 + |U_{\alpha 2}|^2 |U_{\beta 2}|^2 $ and we end up with $(1: 0.504: 0.610)$.
This time assuming both $\nu_3$ and $\nu_2$ decay promptly, we get $\bar{P}_{\alpha\beta}^{\text{inv}} \to |U_{\alpha 1}|^2 |U_{\beta 1}|^2$, and a final flavor ratio of $(1: 0.140: 0.342)$. We can clearly see the relative depletion in the flux of $\nu_\mu$ and $\nu_\tau$ due to the invisible decays.

\subsection{Visible Decay}
\label{ssec:visible decay}
Here we follow and extend upon refs.~\cite{Coloma:2017zpg,Lindner:2001fx}.
The case of visible decay can be qualitatively understood as the depletion of mostly $\nu_\mu$'s and $\nu_\tau$'s, as in invisible decay, accompanied by regeneration consisting of mostly $\nu_e$'s with lower energy in the normal mass ordering.
We now need to account for the regeneration of the neutrino flux with a term describing the appearance of decay products,
\begin{equation}
\bar P_{\alpha\beta}^{\rm vis}(E_f)=\bar P_{\alpha\beta}^{\rm SM}+\bar P_{\alpha\beta}^{\rm dep}(E_f)+\bar P_{\alpha\beta}^{\rm reg}(E_f)\,.
\end{equation}
This section is devoted to calculating this regeneration term which is considerably more complicated, even in the oscillation averaged case, than the SM or depletion terms.
The complexity arises from the fact that the initial and final state neutrino energies, $E_i$ and $E_f$, are now distinct and thus, the observed spectrum depends on the initial spectrum. Additionally, the possibility of consecutive decays must now be accounted for.
Although the effect of multiple decays is smaller than that of single decays, it is not completely negligible.

We first shift to the mass basis since this is the basis in which the decays happen,
\begin{equation}\label{eq:mass_flav}
\bar P^{\text{reg}}_{\alpha\beta}(E_f,L)=\sum_{i>j}|U_{\alpha i}|^2|U_{\beta j}|^2\bar P_{ij}^{\rm reg}(E_f,L)\,.
\end{equation}
The transition probability involves an integral over both the decay location, $L_1$, and the initial neutrino energy, $E_i$,
\begin{multline}
\bar P_{ij}^{\rm reg}(E_f,L)=\\
\frac1{\Phi_i^S(E_f)}
\int_0^LdL_1
\int_{E_f}^{x_{ij}^2E_f}dE_i\Delta\bar P^{\text{reg}}_{ij}(E_i, E_f,L,L_1)\Phi_i^S(E_i)\,,
\label{eq:L1Ei integral}
\end{multline}
where the mass eigenstate spectrum is related to the flavor eigenstate spectrum by $\Phi_i^S=\sum_\alpha|U_{\alpha i}|^2\Phi_\alpha^S$.
In this article we will be assuming that initial spectrum of each flavor is the same up to the overall flavor ratio normalization of $(1:2:0)$.
The integral limits, $E_i\in[E_f,x_{ij}^2E_f]$, are fixed by kinematics due to the allowed final state energies. 

Assuming a neutrino $\nu_i$ decays once at $L_1\le L$, for a baseline $L$, our regeneration term is then constructed from the following five factors:
\begin{enumerate}
    \item the survival of $\nu_i$ over a distance $L_1$, 
    $e^{-\frac{1}{2} \Gamma_i L_1}$,
    \item the phase accumulation of $\nu_i$ from the source to $L_1$, $e^{-iE_iL_1}$,
    \item the decay of $\nu_i$ and appearance of $\nu_j$, $\sqrt{\Gamma_{ij}W_{ij}}$,
    \item for unstable $\nu_j$, survival until Earth,
    $e^{-\frac{1}{2}\Gamma_j(L - L_1)}$, 
    \item and the phase accumulation of $\nu_j$ until Earth,  $e^{-iE_f(L - L_1)}$.
\end{enumerate}
The function $W_{ij} = \frac{1}{\Gamma_{ij}}\frac{d\Gamma\left(E_i,E_f\right)}{dE_f}$ is the normalized energy distribution of the decay products,
\begin{align}
\Gamma_{ij}^{\nu\nu}W_{ij}^{\nu\nu}={}&\frac{m_im_j}{16\pi E_i^2}[g_{ij}^2(A_{ij}+2)+g_{ij}^{\prime2}(A_{ij}-2)]\,,
\label{eq:GW nunu}\\
\Gamma_{ij}^{\nu\bar\nu}W_{ij}^{\nu\bar\nu}={}&\frac{m_im_j}{16\pi E_i^2}(g_{ij}^2+g_{ij}^{\prime2})\left(\frac1{x_{ij}}+x_{ij}-A_{ij}\right)\,,
\label{eq:GW nunubar}\\
\Gamma_{ij}^{\nu\nu,\nu\bar\nu}W_{ij}^{\nu\nu,\nu\bar\nu}={}&\frac{m_im_j}{16\pi E_i^2}\left[g_{ij}^2\left(\frac1{x_{ij}}+x_{ij}+2\right)\right.\nonumber\\
&\left.+g_{ij}^{\prime2}\left(\frac1{x_{ij}}+x_{ij}-2\right)\right]\,,
\label{eq:GW nunu nunubar}
\end{align}
where
\begin{equation}
A_{ij}=\frac1{x_{ij}}\frac{E_i}{E_f}+x_{ij}\frac{E_f}{E_i}\,.
\end{equation}
The superscripts in eq.~\ref{eq:GW nunu nunubar} include both $\nu_i\to\nu_j$ and $\nu_i\to\bar\nu_j$ channels.
To account for specifically helicity conserving (flipping) channels then eqs.~\ref{eq:GW nunu}-\ref{eq:GW nunubar} should be used as appropriate, although the full expression for the total width in eq.~\ref{eq:Gamma} should always be used.

The regeneration amplitude is given by,
\begin{equation}
\mathcal A^{\text{reg}}_{ij}=e^{-\frac{1}{2}\Gamma_i L_1}e^{-\frac{1}{2}\Gamma_j(L - L_1)}\sqrt{\Gamma_{ij}W_{ij}}\,,
\label{eq:Amp}
\end{equation}
where we have removed overall phases that only contribute to oscillations.
The differential decay probability is the amplitude squared,
\begin{equation}
\Delta P_{ij}^{\rm reg}(E_i,E_f,L,L_1)=|\mathcal A_{ij}^{\rm reg}|^2\,.
\label{eq:Delta P reg}
\end{equation}
We use the fact that neutrinos promptly (on scales relative to the total propagation) lose coherency, so we do not include any interference terms. Such terms, however, are important in the context of decays over local distances, and must be accounted for when considering the effects of decay on the measured event rates at, for example, long-baseline neutrino experiments. See appendix \ref{App. ssec. reg} for details.
Averaging over oscillations and assuming all decays happen incoherently, we analytically perform the integral over $L_1$,
\begin{equation}
\int_0^LdL_1\Delta\bar{P}^{\text{reg}}_{ij}(E_i, E_f, L, L_1)=\frac{\Gamma_{ij}W_{ij}}{\Gamma_i - \Gamma_j}\left[1-e^{-(\Gamma_i-\Gamma_j) L}\right]\,.
\end{equation}

Up to this point we have only considered single decays.
If both $g^{(\prime)}_{32}\neq0$ and $g^{(\prime)}_{21}\neq0$, the neutrinos will experience consecutive decays, $\nu_3\to\nu_2\to\nu_1$. 
This additional decay pathway modifies the $\nu_3\to\nu_1$ transition probability. 
The initial neutrino decays at a distance $L_1 \le L$, after which the intermediary neutrino, with energy $E_{\rm int}$, propagates and decays at $L_2$, with $ L_1 \le L_2 \le L$. 
We construct the regeneration amplitude as described above,
\begin{multline}
\mathcal A^{\text{reg},2}_{31}=
e^{-\frac{1}{2}\Gamma_3 L_1}\sqrt{\Gamma_{32}W_{32}}
e^{-\frac{1}{2}\Gamma_2\left(L_2 - L_1\right)} \sqrt{\Gamma_{21}W_{21}}\,,
\end{multline}
noting that $\Gamma_{32}W_{32}=\Gamma_{32}W_{32}(E_i,E_{\rm int})$, $\Gamma_{21}W_{21}=\Gamma_{21}W_{21}(E_{\rm int},E_f)$, and that $\Gamma_3=\Gamma_3(E_i)$, $\Gamma_2=\Gamma_2(E_{\rm int})$. 
The total probability is given by,
\begin{multline}
\bar P_{31}^{\rm reg,2}(E_f,L)=\frac1{\Phi_i^S(E_f)}\\
\times\int_0^LdL_1\int_{L_1}^LdL_2\int_{E_f}^{x_{32}^2 E_f}dE_{\rm int}\int_{E_{\rm int}}^{x_{21}^2 E_{\rm int}}dE_i\\
\times\Delta\bar P^{\text{reg},2}_{31}(E_i,E_{\rm int}, E_f,L,L_1,L_2)\Phi_i^S(E_i)\,.
\end{multline}
We can again perform the $L$ integrals analytically,
\begin{multline}
\int_0^LdL_1\int_{L_1}^LdL_2\Delta\bar{P}^{\text{reg},2}_{31}(E_i, E_{\rm int}, E_f, L_1,L_2)\\
=\frac{\Gamma_{32}W_{32}\Gamma_{21}W_{21}}{\Gamma_3-\Gamma_2}\left[\frac{1}{\Gamma_2}\left(1-e^{-\Gamma_2 L}\right) - \frac{1}{\Gamma_3}\left(1-e^{-\Gamma_3 L}\right)\right]\,,
\label{eq:2decays}
\end{multline}
followed by integration over the initial and intermediate energies.
The $\nu_3\to\nu_1$ transition probability now reads
\begin{equation}
    \bar{P}^{\text{reg}}_{31}\to \bar{P}^{\text{reg}}_{31} + \bar{P}^{\text{reg},2}_{31}\,.
\end{equation}

\subsection{Cosmology}\label{subsec:cosmo}
So far we have assumed that the neutrinos only propagate over local distances. 
In the context of high energy astrophysical neutrinos, it is known that their propagation distances are long enough that the expansion of the universe must be accounted for.
Decay is dependent on the duration of travel, and so light-travel distance is the correct measure for this purpose.
Baselines and neutrino energies are now functions of the redshift to the source, $z$,
\begin{align}
    E(z) &= E_{0}(1+z)\,,\\
    L(z_a,z_b) &= L_H \int_{z_a}^{z_b}  \frac{dz'}{(1+z^{\prime})h(z^{\prime})}\label{eq:dist}\,,
\end{align}
where $L_H=c/H_0$ is the Hubble length, $h(z)\equiv H(z)/H_{0}$ and 
\begin{equation}
    h(z) \equiv \sqrt{\Omega_{m}(1+z)^{3} + \Omega_{\Lambda}}\,.
\end{equation}
The energy $E(z)$ is the energy at production, and $E_0$ the observed energy at detection. The function $L(z_a,z_b)$ computes the distance between two redshift points $z_a$ and $z_b$. 
We assume a flat $\Lambda$CDM cosmology, with $\Omega_{m}=0.315$, $H_0=67.4$ km/s/Mpc \cite{Aghanim:2018eyx} and $\Omega_{\Lambda}= 1-\Omega_{m}$.
With the introduction of cosmology, both the invisible and visible scenarios are more subtle and require some care. In the following section, we will continue to work in the mass basis, but all expressions can be re-expressed in the flavor basis using eq.~\ref{eq:mass_flav}.
Since the neutrino energies are now a function of redshift, the spectrum at the source and at the point of decay are no longer the same, but are modified by a factor,
\begin{align}
\Phi_i^S(E_i)&\to \Phi_i^S(E_i(1+z))\nonumber\\ &=\Phi_i^S(E_i)(1+z)^{-\gamma}\,.
\end{align}
In accordance with the definition given in eq.  \ref{eq:prob_defn}, 
the SM transition probability is also modified by the same factor
\begin{equation}
\bar P^{\rm SM}_{ij}(z)=\delta_{ij}(1+z)^{-\gamma}\,.
\end{equation}
To compute the depletion and regeneration terms, we make the substitutions: $E\to E(1+z)$, which implies $\Gamma_{i}\to\Gamma_{i}/(1+z)$ and $\Gamma_{ij}W_{ij}\to\Gamma_{ij}W_{ij}/(1+z)^2$. As we frequently encounter the product $\Gamma_i L$ in the arguments of our exponential functions, we make a redefinition of the distance integral, eq.~\ref{eq:dist}, 
\begin{equation}
    L(z_a,z_b)\to L(z_a,z_b) = L_H \int_{z_a}^{z_b}  \frac{dz'}{(1+z^{\prime})^2h(z^{\prime})}\label{eq:dist_mod}\,.
\end{equation}
The depletion term is now given by
\begin{equation}
\bar{P}_{ij}^{\text{dep}}\left(E_f,z\right)= -\delta_{ij}(1+z)^{-\gamma}\left(1-e^{-\Gamma_i L(0,z)}\right)\,.
\end{equation}
For the regeneration term, we start from eq.~\ref{eq:Amp} and make the relevant substitutions.
The modified version of eq.~\ref{eq:L1Ei integral} now involves an integral over $z_1$, the redshift at which the neutrino decays,
\begin{multline}
\bar P_{ij}^{\rm reg}(E_f,z)=\frac{1}{\Phi_i^S(E_f)}\\
\times\int_0^z \frac{dL}{dz_1}dz_1\int_{E_f(1+z_1)}^{x_{ij}^2E_f(1+z_1)}(1+z_1)dE_i\\
\times\Delta\bar P^{\text{reg}}_{ij}(E_i, E_f,z_1)\Phi_i^S(E_i(1+z_1))\,,
\label{eq:z1Ei integral}
\end{multline}
where 
\begin{equation}
\frac{dL}{dz} = -\frac{L_H}{(1+z)h(z)}\,.
\end{equation}
After making the correct substitutions and taking the squared amplitude, we obtain
\begin{multline}
\bar P_{ij}^{\rm reg}(E_f,z)=\frac{L_H}{\Phi_i^S(E_f)}
\int_0^z dz_1 
\int_{E_f(1+z_1)}^{x_{ij}^2E_f(1+z_1)}dE_i\\
\times \frac{\Gamma_{ij}W_{ij} e^{-\Gamma_i L(z_1,z)-\Gamma_j L(0,z_1)}}{(1+z_1)^2h(z_1)}\Phi_i^S(E_i(1+z))\,.
\label{eq:Preg1 cosmo complete}
\end{multline}

In the case of consecutive decays, as in sec.~\ref{ssec:visible decay}, we account for integration over the redshift at which our second decay occurs by adding a second term for $\nu_3\to\nu_1$,
\begin{equation}
\bar P^{\text{reg}}_{31} \to  \bar P^{\text{reg}}_{31} + \bar P^{\rm reg,2}_{31}\,,
\end{equation}
where the first term is from eq.~\ref{eq:Preg1 cosmo complete} and the second is,
\begin{widetext}
\begin{equation}
\bar P^{\rm reg,2}_{31}=\frac{L_H^2}{\Phi_i^S(E_f)}
\int_{0}^{z}dz_2\int_{z_2}^{z}dz_1\int_{E_f(1+z_2)}^{E_fx_{21}^2(1+z_2)}dE_{\rm int}\int_{E_{\rm int}(1+z_1)}^{E_{\rm int}x_{32}^2(1+z_1)}dE_i
\frac{\Gamma_{32}W_{32}\Gamma_{21}W_{21}e^{-\Gamma_3L(z_1,z)-\Gamma_2L(z_2,z_1)}}{(1+z_1)^2h(z_1)(1+z_2)^2h(z_2)}\Phi_i^S(E_i(1+z_1))\,,
\end{equation}
\end{widetext}
where we note that
\begin{align}
\Gamma_{32}W_{32}&\to\Gamma_{32}W_{32}(E_i,E_{\rm int})\,,\\
\Gamma_{21}W_{21}&\to\Gamma_{21}W_{21}(E_{\rm int},E_f)\,.
\end{align}

Finally, since not all neutrinos are coming from the same redshift, we must integrate the total (in)visible probability over the redshift distribution of the source population $R(z)$,
\begin{equation}
\bar P_{\alpha\beta}^{\rm vis/inv}(E_f)=\frac{\int_0^{z_{\max}}dz\bar P_{\alpha\beta}^{\rm vis/inv}(E_f,z)R(z)}{\int_0^{z_{\max}}dzR(z)}\,,
\end{equation}
where we note that we must include depletion and regeneration components as appropriate (the SM part pulls out of the integral since it does not depend on redshift).
For simplicity we take $R(z)=\delta(z-1)$ as our benchmark redshift evolution as this roughly reproduces typical redshift evolution functions, see below.

\section{Analytic Solutions}
\label{sec:analytic}
While a closed form solution for the regeneration term including cosmology likely does not exist, it is possible to find useful expressions without cosmology.
These provide intuition for what is going on and are relevant, for example, for neutrinos propagating over galactic distances such as from a galactic supernova wherein cosmology does not play a role.
For simplicity we consider only one of the $g^{(\prime)}_{ij}\neq0$.
We carry out the integrals discussed in \ref{ssec:visible decay} including a power law spectrum and find,
\begin{multline}
\bar P^{\rm reg}_{ij}(E_f,L)=\frac{z(x)}{\gamma y(x)}\left\{1-\frac1{x^{2\gamma}}\right.\\
\left.+\gamma T^{-\gamma}
\left[\Gamma(\gamma,T)-\Gamma\left(\gamma,\frac T{x^2}\right)\right]\right\}\,.
\label{eq:analytic}
\end{multline}
where $\Gamma(a,x)\equiv\int_x^\infty t^{a-1}e^{-t}dt$ is the incomplete gamma function, $x\equiv x_{ij}=m_i/m_j$, and
\begin{align}
T&=\frac{m_im_jL}{16\pi E_f}y(x)\,,\\
y(x)&=g_{ij}^2\left[f(x)+k(x)\right]+g^{\prime2}_{ij}\left[h(x)+k(x)\right]\,.\\
z(x)&=g_{ij}^2\left(\frac1x+x+2\right)+g_{ij}^{\prime2}\left(\frac1x+x-2\right)\,.
\end{align}
We validate this expression in appendix \ref{sec:analytic validate}.
While eq.~\ref{eq:analytic} is relatively intractable, it does allow for several interesting limits to be evaluated.

\subsubsection*{Limiting Cases}
First, we confirm that as $E_f\to\infty$ (which takes us back to the SM) the regeneration probability goes to zero.
To see this we use the fact that
\begin{equation}
\lim_{T\to0}T^{-\gamma}\left[\Gamma\left(\gamma,T\right)-\Gamma\left(\gamma,\frac T{x^2}\right)\right]=-\frac1\gamma\left(1-\frac1{x^{2\gamma}}\right)\,,
\end{equation}
which when inserted into eq.~\ref{eq:analytic}, gives zero.

Next, as $E_f\to0$ (the full decay scenario, $\Gamma_i\to\infty$) the regeneration probability tends to a non-zero constant.
In this limit $T\to\infty$ and thus $T^{-\gamma}\Gamma(\gamma,T)\to0$, so
\begin{equation}
\lim_{E_f\to0}\bar P^{\rm reg}_{ij}(E_f,L)=\frac{z(x)}{\gamma y(x)}\left(1-\frac1{x^{2\gamma}}\right)\,.
\end{equation}

Within this limit, we consider the heavy mass case, $m_1\to\infty$ ($x\to1$) and find
\begin{equation}
\lim_{\substack{\mathllap{E_f}\to\mathrlap{0}\\\mathllap{m_1}\to\mathrlap{\infty}}}\bar P_{ij}^{\rm reg}(E_f,L)=1\,.
\end{equation}
Taking only $g^{(\prime)}_{31}\neq0$, without cosmology and as $E_f\to0$ and $m_1\to\infty$,
\begin{equation}
\bar P_{\alpha\beta}^{\rm vis}=|U_{\alpha1}|^2|U_{\beta1}|^2+|U_{\alpha2}|^2|U_{\beta2}|^2+|U_{\alpha3}|^2|U_{\beta1}|^2\,,
\end{equation}
and similar for the other $g^{(\prime)}_{ij}$.
It is noteworthy that in this limit the probability is no longer dependent on $\gamma$. This makes sense as the integral over the spectrum is over a region $[E_f,x^2E_f]$. As $x\to1$, the spectrum is asymptotically constant and the final neutrino has the same energy as the initial neutrino.

Finally, taking the small mass case of the low energy limit, $m_1\to0$ ($x\to\infty$)
\begin{equation}
\lim_{\substack{\mathllap{E_f}\to\mathrlap{0}\\\mathllap{m_1}\to\mathrlap{0}}}\bar P_{ij}^{\rm reg}(E_f,L)=\frac1\gamma\,.
\end{equation}
Unlike the previous case, we have dependence on the spectral index $\gamma$.
As expected, a steeper spectrum at the source depresses the low energy regeneration.
That is, if only $g^{(\prime)}_{31}\neq0$, without cosmology and as $E_f\to0$ and $m_1\to 0$,
\begin{equation}
\bar P_{\alpha\beta}^{\rm vis}=|U_{\alpha1}|^2|U_{\beta1}|^2+|U_{\alpha2}|^2|U_{\beta2}|^2+\frac{|U_{\alpha3}|^2|U_{\beta1}|^2}\gamma\,.
\end{equation}
We also note that in this limit the probability is independent of $m_1$ when $\gamma=1$.
These results recover the same expressions as were found in ref.~\cite{Beacom:2002vi}.

Some closed form solutions exist for arbitrary $E_f$ and $m_1$ that do not depend on incomplete gamma functions.
In our benchmark case of $\gamma=2$, we can use the fact that $\Gamma(2,x)=e^{-x}(1+x)$ to write
\begin{multline}
\bar P^{\rm reg}_{ij}(E_f,L)=\frac{z(x)}{2y(x)}\left\{1-\frac1{x^4}\right.\\
\left.+\frac2{T^2}\left[e^{-T}(1+T)-e^{-T/x^2}\left(1+\frac T{x^2}\right)\right]\right\}\,.
\end{multline}
Additional closed form solutions exist for other integer values of $\gamma$; e.g.~$\Gamma(3,x)=e^{-x}(x^2+2x+2)$ and so on.

While a full expression for the regeneration term with cosmology is extremely complicated, the various limits evaluate quite simply as for the case without cosmology.
We find that, including cosmology, in the heavy mass case
\begin{equation}
\lim_{\substack{\mathllap{E_f}\to\mathrlap{0}\\\mathllap{m_1}\to\mathrlap{\infty}}}\bar P_{ij}^{\rm reg}(E_f,L)=(1+z)^{-2\gamma}\,,
\end{equation}
which if only $g^{(\prime)}_{31}\neq0$ gives
\begin{multline}
\lim_{\substack{\mathllap{E_f}\to\mathrlap{0}\\\mathllap{m_1}\to\mathrlap{\infty}}}\bar P_{\alpha\beta}^{\rm vis}=(1+z)^{-\gamma}\bigg[|U_{\alpha1}|^2|U_{\beta1}|^2+|U_{\alpha2}|^2|U_{\beta2}|^2 \\
+ (1+z)^{-\gamma}|U_{\alpha3}|^2|U_{\beta1}|^2\bigg]\,.
\label{eq:Pvis cosmo Ef0 m1infinity}
\end{multline}
In the light mass case we have
\begin{equation}
\lim_{\substack{\mathllap{E_f}\to\mathrlap{0}\\\mathllap{m_1}\to\mathrlap{0}}}\bar P_{ij}^{\rm reg}(E_f,L)=\frac{(1+z)^{-2\gamma}}\gamma\,,
\end{equation}
and, if again only $g^{(\prime)}_{31}\neq0$, gives
\begin{multline}
\lim_{\substack{\mathllap{E_f}\to\mathrlap{0}\\\mathllap{m_1}\to\mathrlap{0}}}\bar P_{\alpha\beta}^{\rm vis}=(1+z)^{-\gamma}\bigg[|U_{\alpha1}|^2|U_{\beta1}|^2+|U_{\alpha2}|^2|U_{\beta2}|^2\\
+\frac{(1+z)^{-\gamma}}{\gamma}|U_{\alpha3}|^2|U_{\beta1}|^2\bigg]\,.
\label{eq:Pvis cosmo Ef0 m10}
\end{multline}
One factor of $(1+z)^{-\gamma}$ comes from the definition of the probability in eq.~\ref{eq:prob_defn} and the other from the contribution due to cosmology.
This allows us to easily write down the flavor ratios in the full decay limit.
The flavor ratios from pion decay in the full decay case with cosmology and for $m_1=0$ is shown in table \ref{tab:flavor ratios} for various spectral indices in the case of only $\nu_3\to\nu_1$ decay. The flavor ratios in the SM and for invisible decay do not change subject to cosmology. 
We see that as $\gamma\to\infty$ we recover the invisible decay case regardless of $m_1$.
This is because the regeneration term goes to zero relative to the SM and the depletion terms, as shown in eqs.~\ref{eq:Pvis cosmo Ef0 m1infinity} and \ref{eq:Pvis cosmo Ef0 m10}.

To summarize, in the full decay limit, there are four main cases as discussed above: $m_1\to0$ or $m_1\to\infty$ and with or without cosmology.
Only in the case of $m_1\to\infty$ and without cosmology does the spectral index not affect the final flavor ratio.

\begin{table}
\centering
\caption{The flavor ratios for full decay with cosmology for various spectral indices at the benchmark point with $g^{(\prime)}_{21}=g^{(\prime)}_{32}=0$.
In the SM and invisible cases, the flavor ratios are $(1:1.152:1.117)$ and $(1:0.504:0.610)$ respectively.}
\begin{tabular}{|c|c|}
\hline
$\gamma$&Flavor ratio\\\hline
1&(1:0.396:0.531)\\\hline
2&(1:0.469:0.584)\\\hline
3&(1:0.492:0.601)\\\hline
4&(1:0.499:0.606)\\\hline
5&(1:0.502:0.608)\\\hline
\end{tabular}
\label{tab:flavor ratios}
\end{table}

\begin{figure}
\centering
\includegraphics[width=\columnwidth]{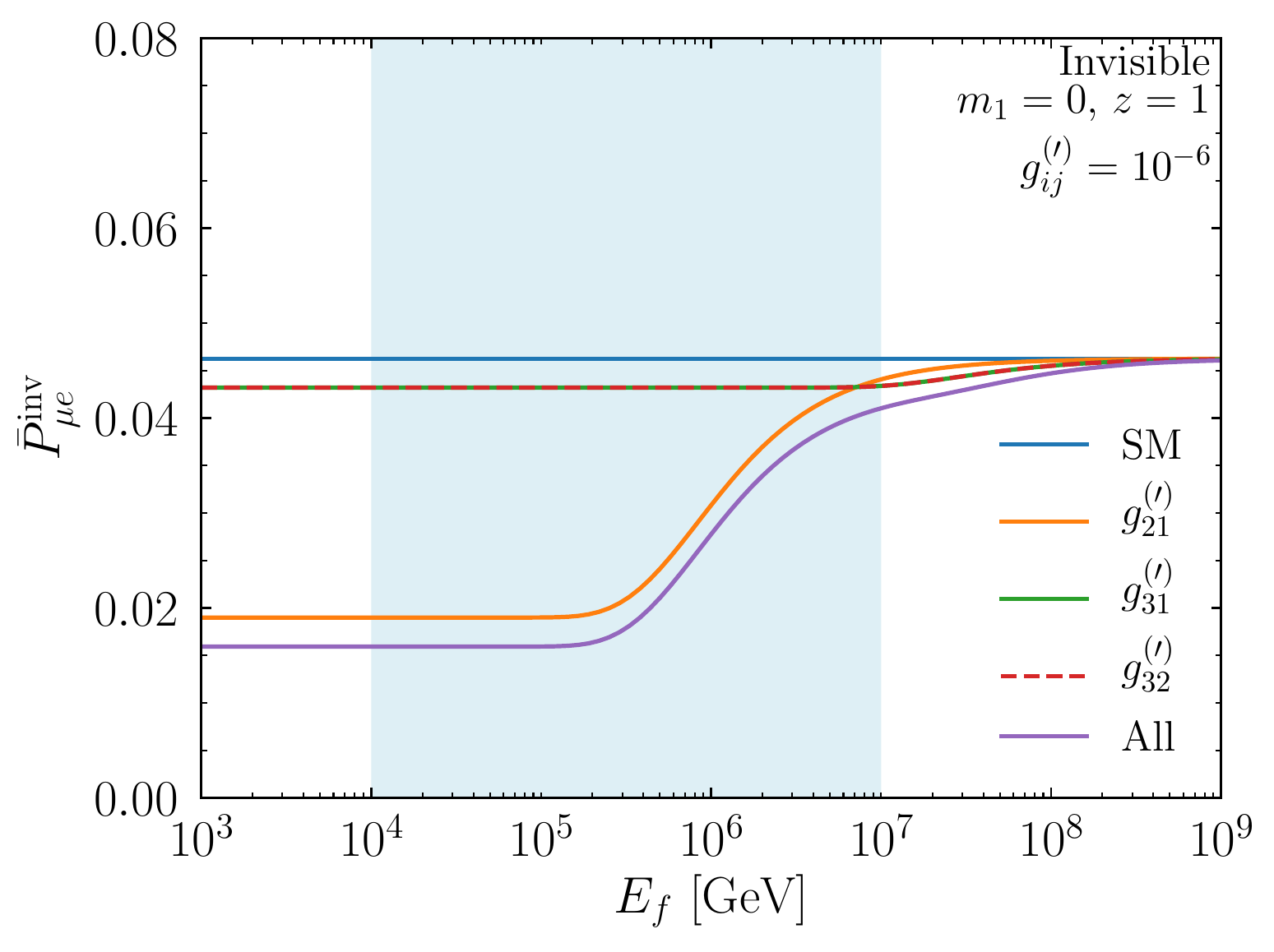}
\caption{The oscillation averaged invisible decay probability for benchmark parameters (see table \ref{tab:benchmark}) including both scalar and pseudo-scalar interactions and decays to both neutrinos and antineutrinos.
The orange, green, and red curves are all for one channel at a time with $g_{ij}=g'_{ij}$, and the purple curve is for all six couplings non-zero.
The $(31)$ and $(32)$ cases only slightly differ.
The light blue region is the widest energy range that IceCube is likely to be sensitive to.}
\label{fig:invisible}
\end{figure}

\section{Results}
\label{sec:results}
We now numerically integrate the expressions in section \ref{sec:formalism}.
As there are numerous parameters in our model, we consider benchmark parameters shown in table \ref{tab:benchmark} and then numerically show how deviations from the benchmark parameters affect the transition probability.
We take all six couplings to be $10^{-6}$, including both scalar and pseudo-scalar couplings; this puts the decay features in IceCube's region of interest.
We take the lightest neutrino to be massless, the initial spectral index to be 2, and the redshift evolution to be a delta function at $z=1$.
We consider both $\nu\to\nu$ and $\nu\to\bar\nu$ channels as is relevant for Majorana neutrinos.
To better illustrate the effect of our parameters on the probability, we select a single channel, $\nu_\mu\to\nu_e$, to focus on and show the rest in fig.~\ref{fig:visible f}.

\begin{figure*}
\centering
\includegraphics[width=0.49\textwidth]{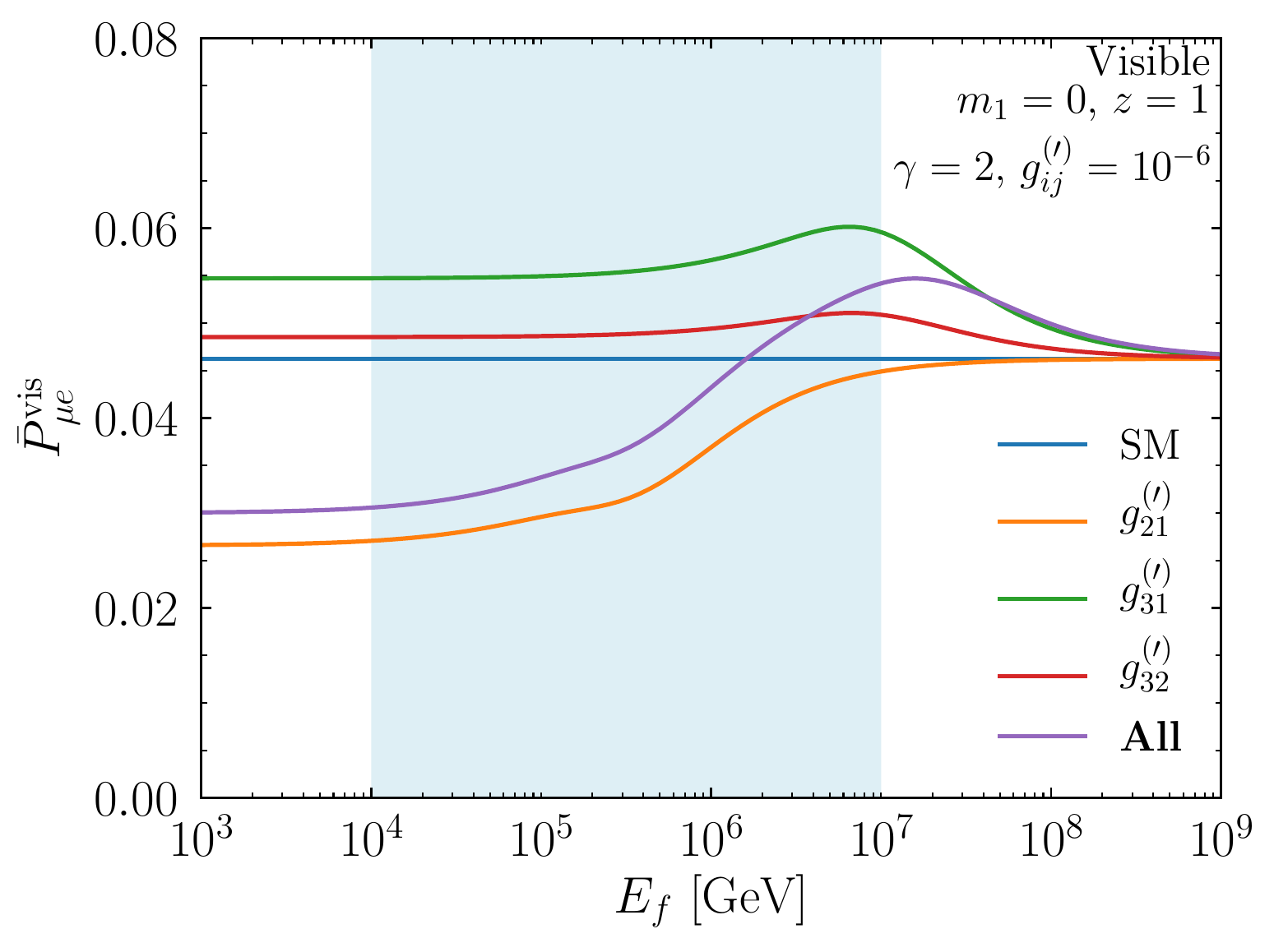}
\includegraphics[width=0.49\textwidth]{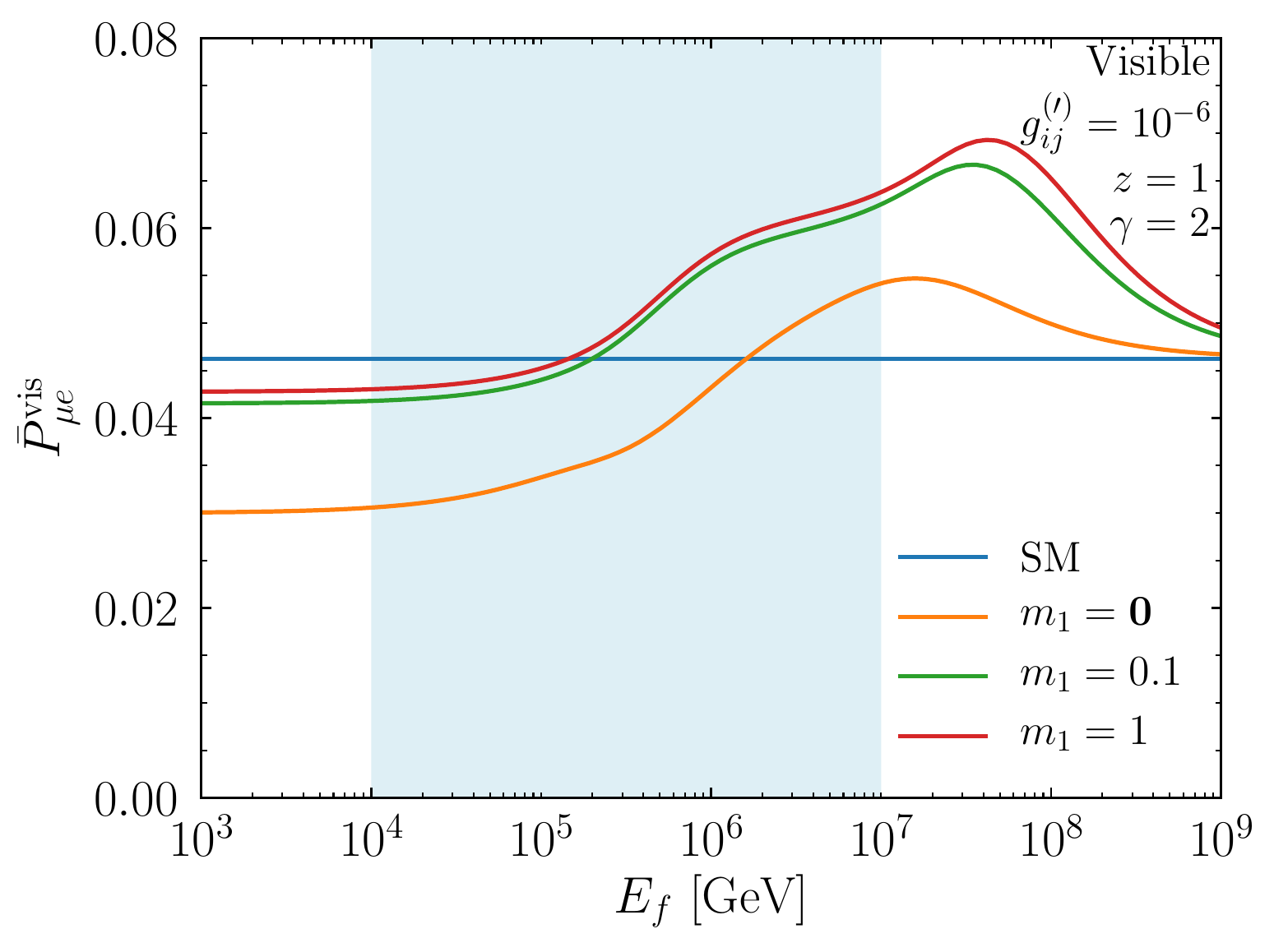}
\includegraphics[width=0.49\textwidth]{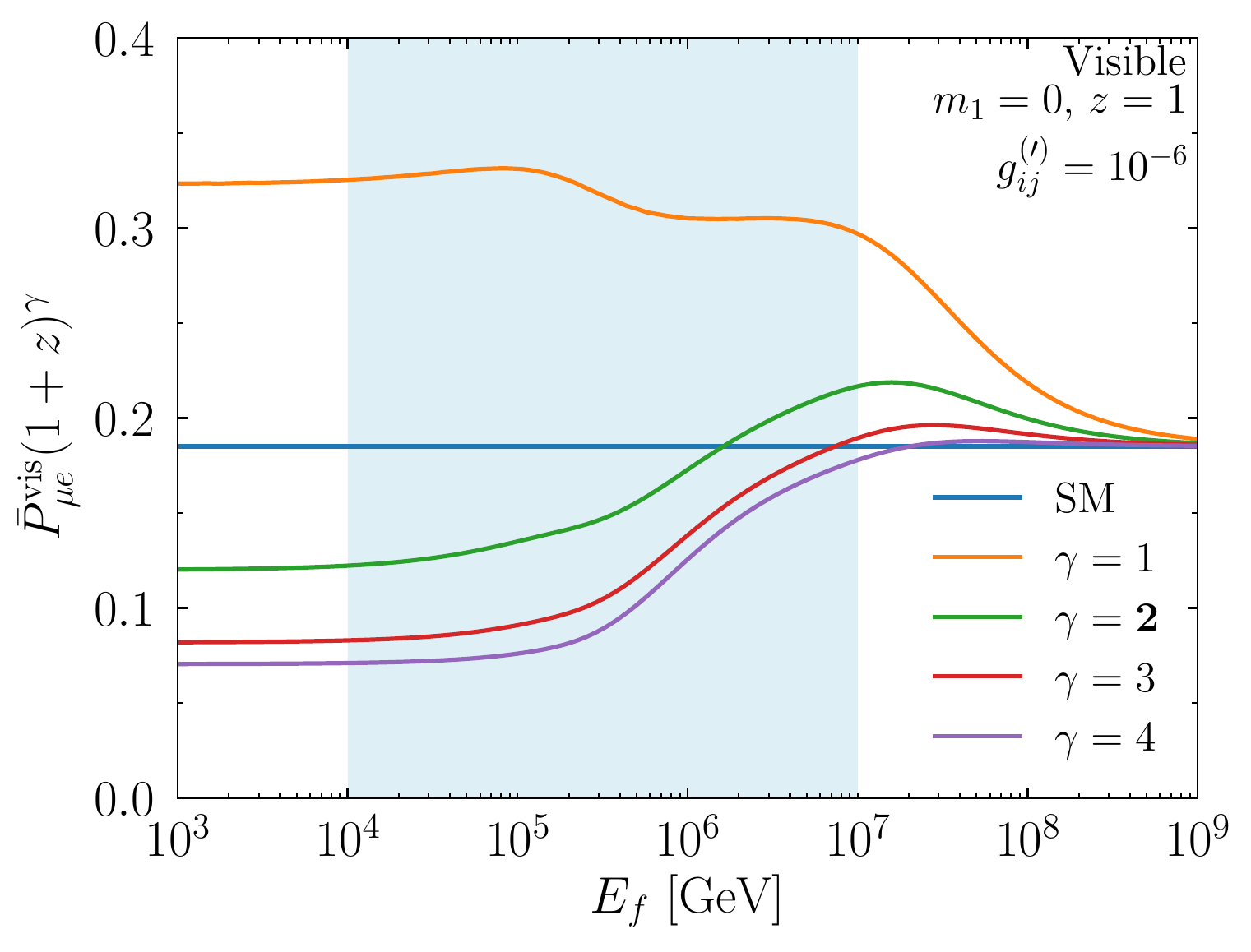}
\includegraphics[width=0.49\textwidth]{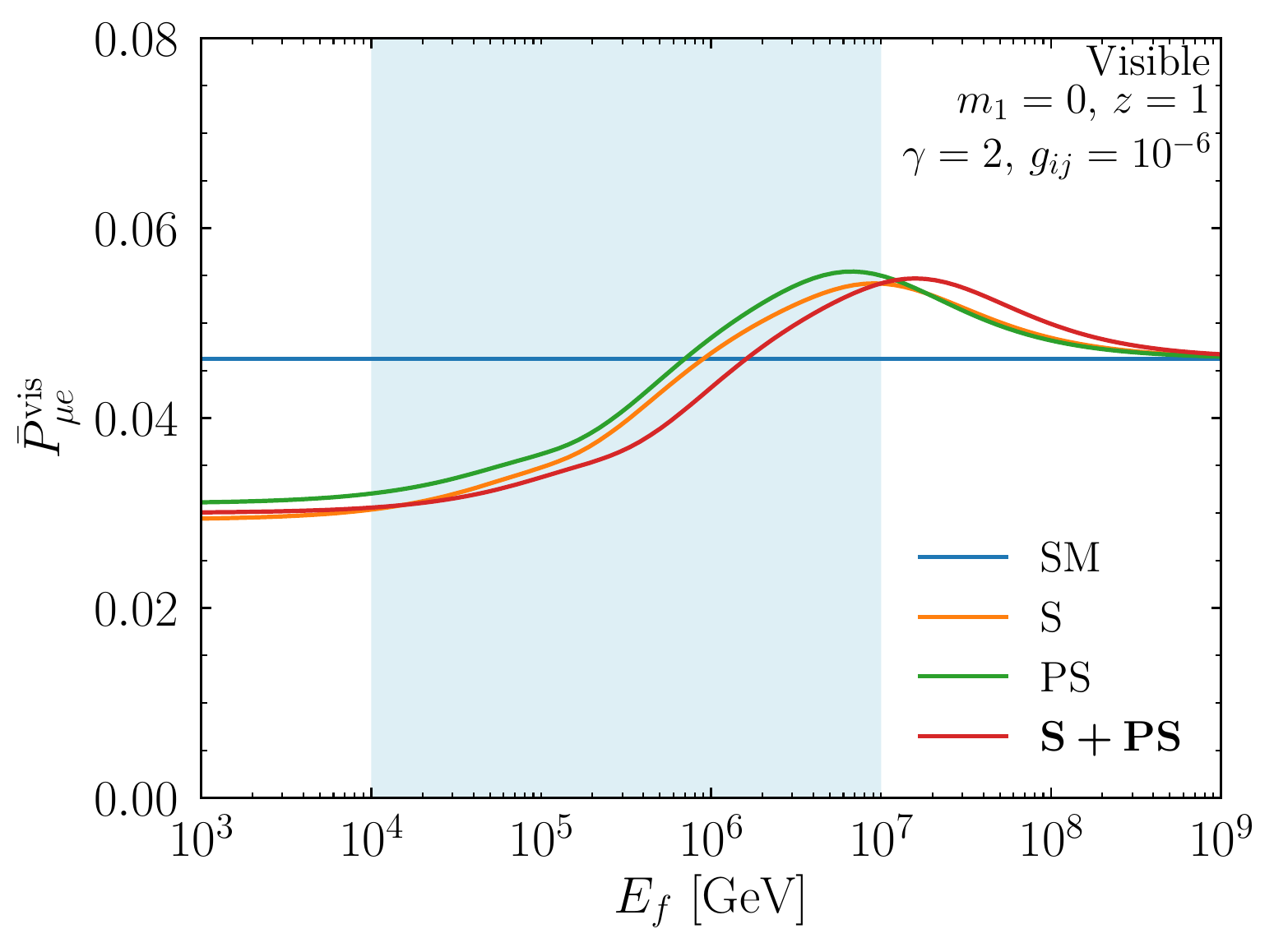}
\includegraphics[width=0.49\textwidth]{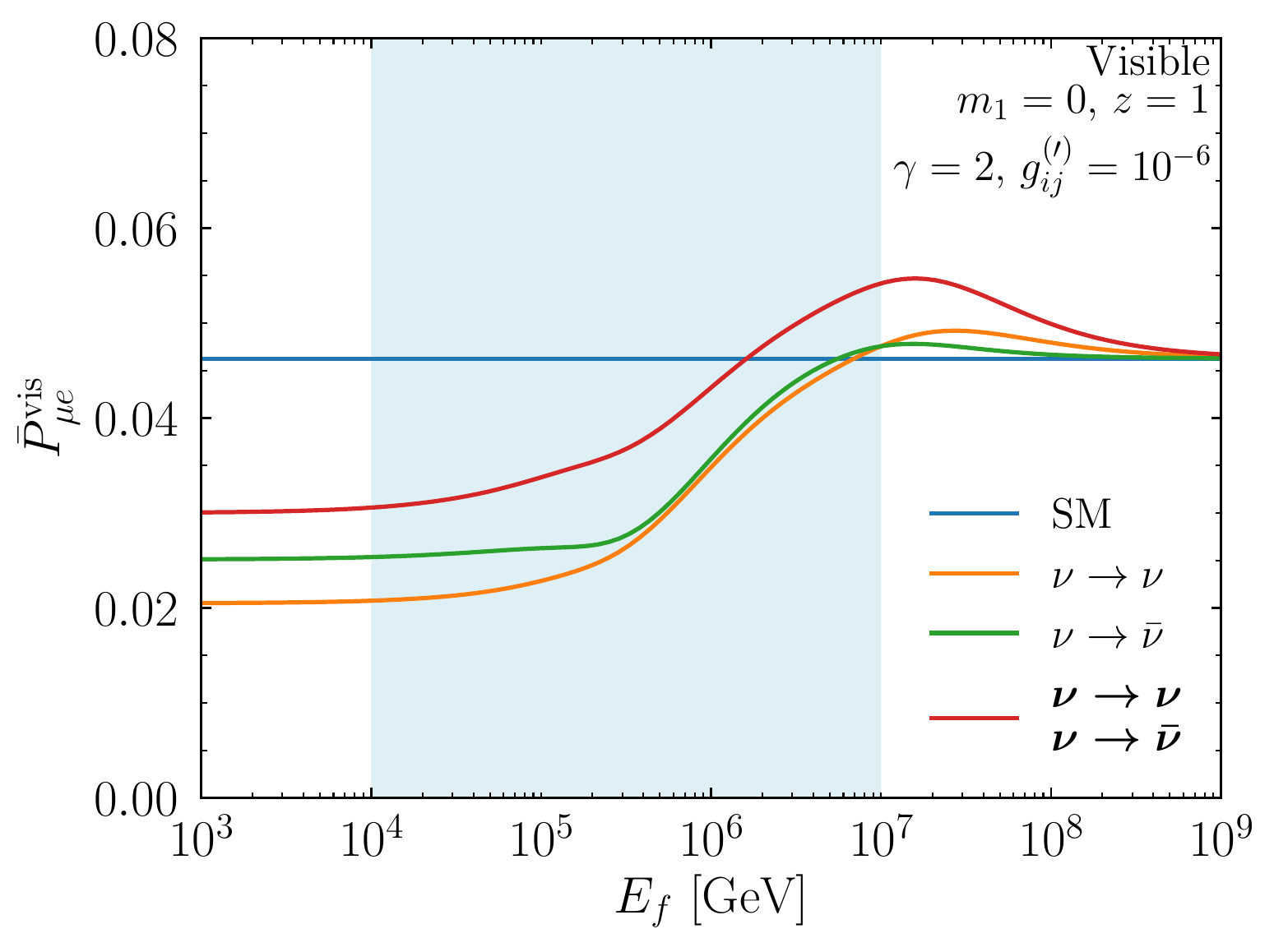}
\includegraphics[width=0.49\textwidth]{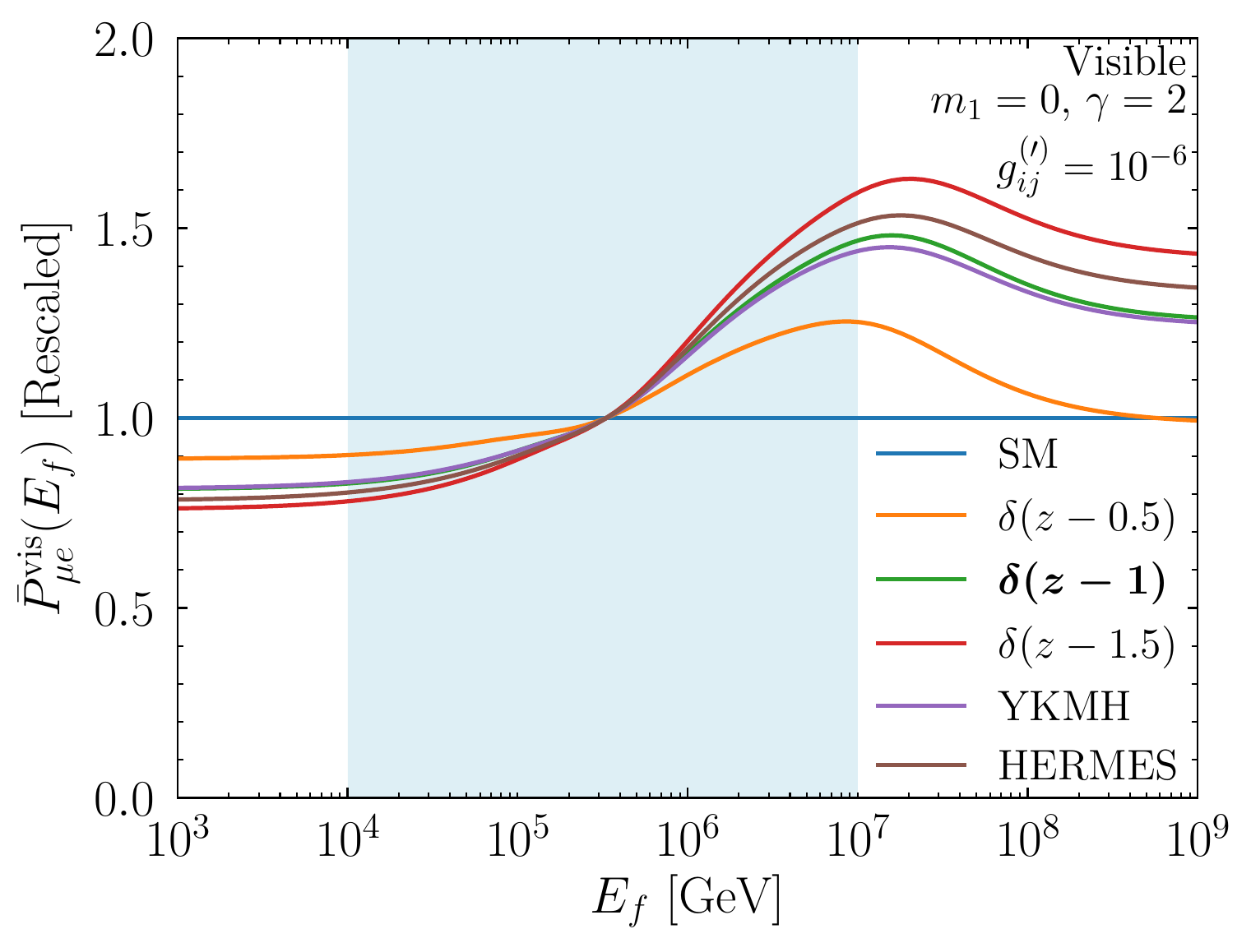}
\caption{The $\nu_\mu\to\nu_e$ oscillation averaged probability traveling to the Earth including cosmology for the benchmark visible decay parameters (see table \ref{tab:benchmark}) except where otherwise specified.
\textbf{Top Left}: Varying which couplings are on with $g_{ij}=g'_{ij}$.
\textbf{Top Right}: Varying lightest neutrino mass $m_1$ in eV.
\textbf{Middle Left}: Varying the source spectral index $\gamma$ where all curves are rescaled by $(1+z)^\gamma$ so they asymptote to the same value.
\textbf{Middle Right}: Scalar interactions ($g_{ij}$ terms only), pseudo-scalar interactions ($g'_{ij}$ terms only), or both $g_{ij}$ and $g'_{ij}$ terms.
\textbf{Bottom Left}: Varying neutrino and antineutrino decay channels.
The red curve is the Majorana case and the orange and green curves are the Dirac case depending on the initial $\nu/\bar\nu$ ratio.
\textbf{Bottom Right}: Varying the redshift evolution where all curves are normalized to $E_f\simeq300$ TeV for convenience.}
\label{fig:visible}
\end{figure*}

\begin{figure}
\centering
\includegraphics[width=\columnwidth]{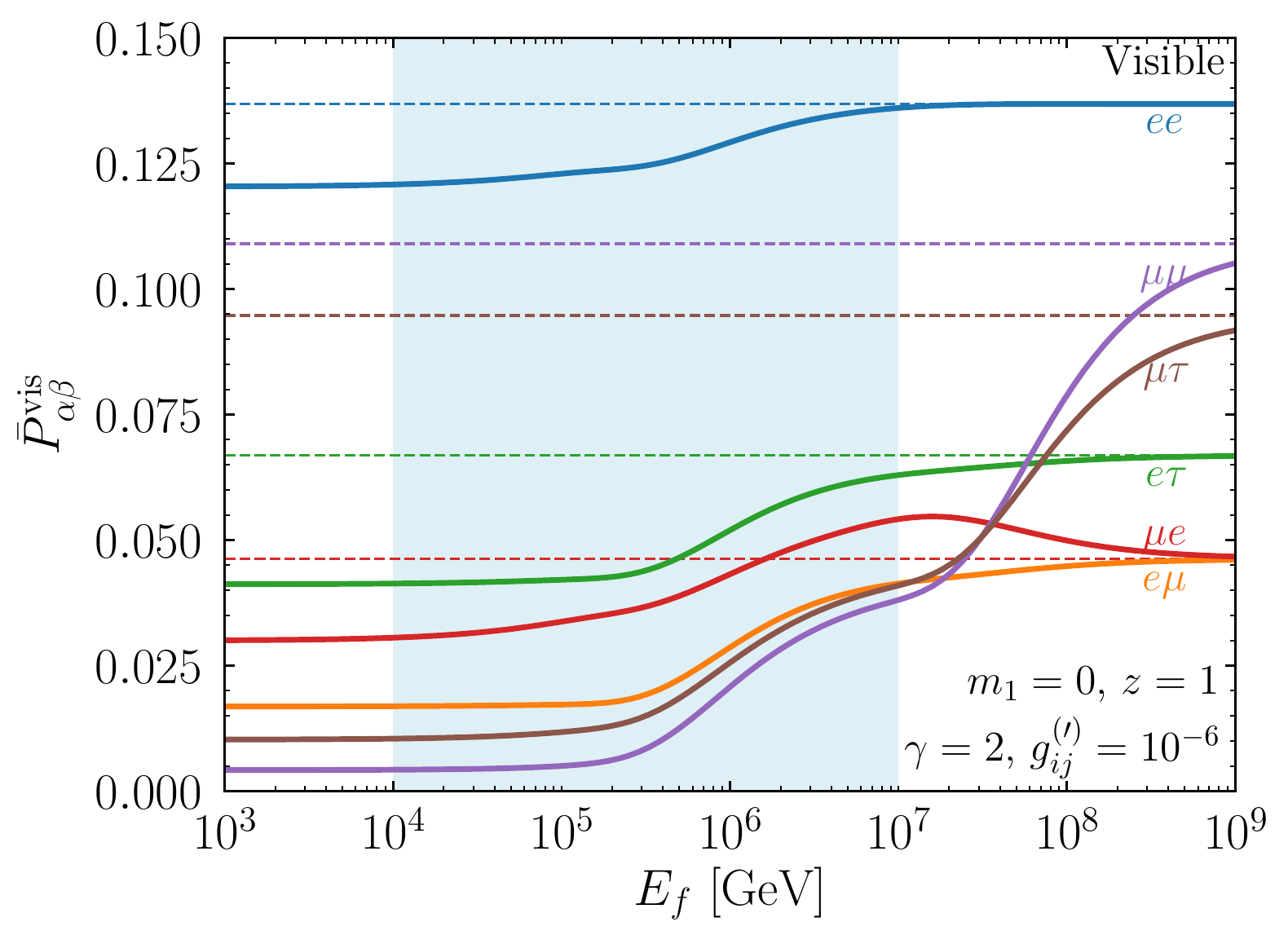}
\caption{The same as in fig.~\ref{fig:visible} but now for different channels.
The dashed curves show the SM case for each channel (note that $\bar P_{\mu e}^{\rm SM}=\bar P_{e\mu}^{\rm SM}$.}
\label{fig:visible f}
\end{figure}

In fig.~\ref{fig:invisible} we show the oscillation averaged probability for $\nu_\mu\to\nu_e$ in the case of invisible decay.
This shows that each specific flavors will receive a unique energy dependent modification that also depends on the structure of the coupling matrices, $g_{ij}$ and $g'_{ij}$.
In fig.~\ref{fig:visible}, we show the visible decay case for the benchmark point, and then show the effect of varying which coupling is turned on, the absolute neutrino mass scale, the initial spectrum, and the redshift evolution of the sources.
Varying each parameter yields a unique modification of the probability and, in principle, can be identified up to experimental precision and astrophysical uncertainties.
For the redshift evolution, in addition to $R(z)=\delta(z-1)$, we also consider two from various fits to astrophysical objects which could potentially be sources.
We refer to the first one as HERMES \cite{Gruppioni:2013jna},
\begin{equation}
R(z)=
\begin{cases}
(1+z)^m\quad&z<z_c\,,\\
(1+z_c)^m&z>z_c\,,
\end{cases}
\end{equation}
with $m\simeq3$ and $z_c\simeq1.5$, and the second as YKBH \cite{Yuksel:2008cu},
\begin{equation}
R(z)=\left[(1+z)^{p_1k}+\left(\frac{1+z}{5000}\right)^{p_2k}+\left(\frac{1+z}9\right)^{p_3k}\right]^{1/k}\,,
\end{equation}
with $p_1=3.4$, $p_2=-0.3$, $p_3=-3.5$, and $k=10$ up to $z_{\max}=5$.
We see that $R(z)=\delta(z-1)$ fits inbetween these two redshift evolution functions in the bottom right panel of fig.~\ref{fig:visible}, justifying its use as a benchmark parameter; it is considerably simpler computationally.

\begin{table}
\centering
\caption{The benchmark parameters used unless otherwise specified.
S,PS refers to the presence of scalar or pseudo scalar interactions; both are included by default.
The effects of varying these parameters are shown in fig.~\ref{fig:visible}.}
\begin{tabular}{|c|c|c|c|c|c|c|c|}
\hline
$g^{(\prime)}_{21}$&$g^{(\prime)}_{31}$&$g^{(\prime)}_{32}$&$m_1$&$\gamma$&S,PS&$R(z)$&$\nu$,$\bar\nu$\\\hline
$10^{-6}$&$10^{-6}$&$10^{-6}$&0 eV&2&S+PS&$\delta(z-1)$&$\nu\to\nu$, $\nu\to\bar\nu$\\\hline
\end{tabular}
\label{tab:benchmark}
\end{table}

The $\nu_\mu\to\nu_e$ channel is not the only relevant channel for astrophysical neutrino decay; there are nine channels governing neutrino decay in total, but six are relevant in our context.
Since no $\nu_\tau$'s are produced in sources, the $\nu_\tau\to\nu_\alpha$ channels are not relevant.
In addition, we note that even though this is an oscillation averaged calculation and we have assumed a loss of coherency throughout (hence no CP violating interference terms), $P_{\alpha\beta}\neq P_{\beta\alpha}$ for $\alpha\neq\beta$ except at high energies (that is, the SM). This is, of course, because while $\nu_3\to\nu_1$, $\nu_3\to\nu_2$ and $\nu_2\to\nu_1$ decays may proceed, the reverse are not kinematically allowed. There are thus six main channels of interest.
Fig.~\ref{fig:visible f} shows the probabilities for these six channels in the SM and for our benchmark visible decay scenario.
In this figure we can see that $P_{\mu e}>P_{e\mu}$ for all energies as expected in the normal mass ordering where a regeneration of electron neutrinos is expected to be larger than a regeneration of muon neutrinos since electron neutrinos contain more $\nu_1$ than muon neutrinos do.

\section{IceCube}
\label{sec:icecube}
IceCube is sensitive to the high energy astrophysical neutrino flux over a range of energies spanning from $\sim10$ TeV to $\sim10$ PeV, although most of their sensitivity is between 100 TeV and 1 PeV.
Below $\sim100$ TeV the backgrounds from atmospheric neutrinos and muons become dominant and above 1 PeV the flux has fallen off to just a few events thus far.

IceCube has some sensitivity to the flavor of the neutrino, primarily through the identification of track and cascade topologies.
If a neutrino interacts via a charged current (CC) interaction as a $\nu_\mu$, a muon will be produced leaving a long track.
On the other hand, a $\nu_e$ or $\nu_\tau$ CC interaction will result in an electromagnetic or hadronic shower, making them very hard to differentiate, although in principle possible \cite{Li:2016kra}.
However, neutral current (NC) neutrino interactions also result in hadronic showers, although with $\sim1/3$ of the total neutrino energy.
It is also possible for a $\nu_\tau$ CC interaction to result in a track if the tau decays to a muon, although this branching ratio is 17.4\% \cite{Tanabashi:2018oca} and the resulting muon carries about $\sim1/3$ the energy as it would from a $\nu_\mu$ CC interaction.
Finally, tracks and cascade topologies are sometimes misidentified \cite{Aartsen:2015ivb}.
The corrections due NC interactions and misidentification are very small and so, to an excellent approximation, the track flux is just the $\nu_\mu$ flux and the cascade flux the sum of the $\nu_e$ and $\nu_\tau$ fluxes \cite{Denton:2018aml}.

In addition, IceCube has some, albeit very limited, sensitivity in identifying $\nu_\tau$ CC interactions by the spatial or temporal separation between the initial hadronic shower and the subsequent tau decay, although only 1-2 events have been detected using these methods so far \cite{Aartsen:2015dlt,Aartsen:2017mau,Usner:2017aio,Stachurska:2019wfb}.

An experimentally motivated useful quantity to consider is the track to cascade ratio which we approximate by,
\begin{equation}
R_{\rm tc}(E_f)\equiv\frac{\Phi_{\nu_\mu}(E_f)}{\Phi_{\nu_e}(E_f)+\Phi_{\nu_\tau}(E_f)}\,.
\end{equation}
There are some corrections to this as mentioned above, but their effects are very small primarily due to the steeply falling spectrum \cite{Denton:2018aml}.

\begin{figure}
\centering
\includegraphics[width=\columnwidth]{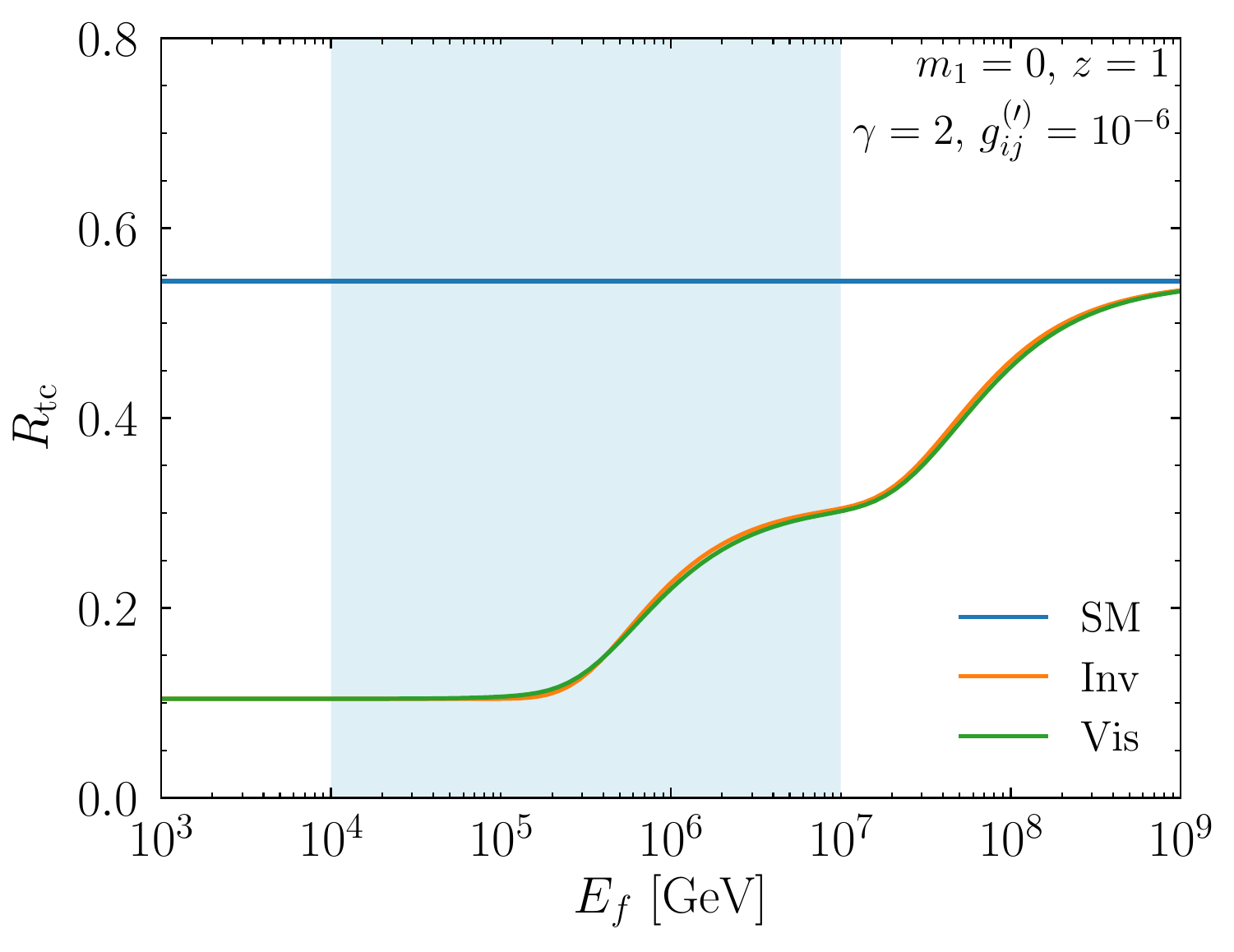}
\caption{The track to cascade ratio as a function of neutrino energy at the Earth for the SM, invisible decay, and visible decay for the benchmark parameters assuming an initial flavor ratio $(1:2:0)$.}
\label{fig:rtc}
\end{figure}

In the SM, this ratio $R_{\rm tc}$ should be constant in energy and independent of the source spectrum or the number of source classes\footnote{The only exception to this is in the case of damped muon decay where the source is dense enough that high energy pions can be produced and decay, but the muons produced from pion decay lose a significant amount of energy before decaying.
We do not consider this case here because a) it will only happen within IceCube's region of interest for a fairly narrow range of astrophysical parameters and b) it yields a fairly small effect ($\Delta\gamma_f\lesssim0.2$) at the Earth after oscillations even for optimal parameters \cite{Denton:2018aml}.}.
The track to cascade ratio, $R_{\rm tc}$, is shown in fig.~\ref{fig:rtc} for the benchmark parameters assuming full pion decay $(1:2:0)$ at the source which oscillates/decoheres to $\sim\frac12$ without neutrino decay.
While both the visible and invisible cases are quite similar, specific structure in $g_{ij}$ can change this such as if only $g^{(\prime)}_{21}$ is non-zero.

\begin{figure}
\centering
\includegraphics[width=\columnwidth]{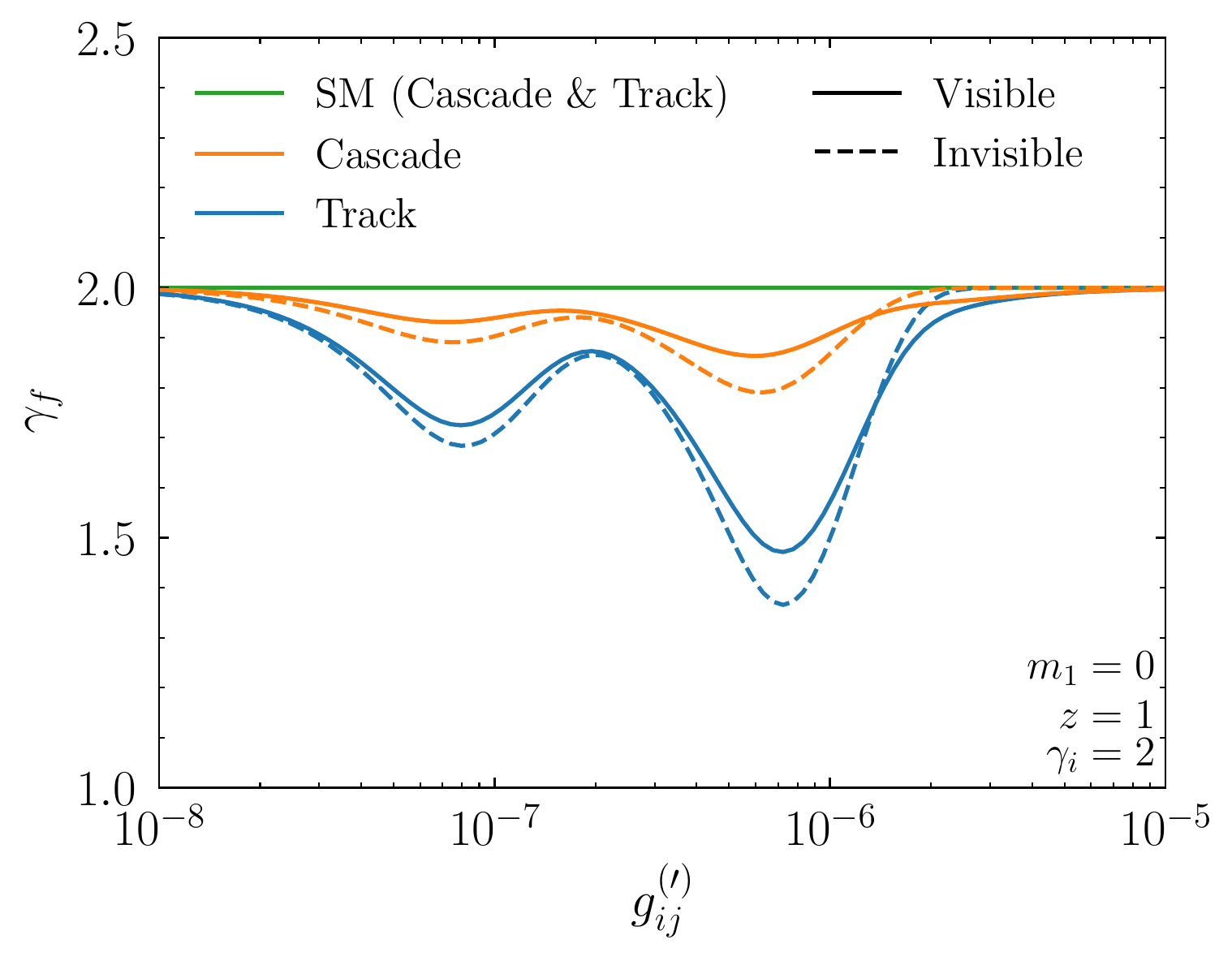}
\caption{The inferred final spectral index at the Earth as a function of coupling for benchmark parameters where we keep all three couplings equal.
In the SM both the track and cascade spectra are the same as the initial spectra, while neutrino decay modifies each spectrum separately assuming an initial flavor ratio $(1:2:0)$.}
\label{fig:IC gamma}
\end{figure}

While neutrino decay leads to a non-trivial spectrum in each flavor, since IceCube fits a power law to their spectrum, we follow the same procedure to determine the effect of neutrino decay at IceCube.
We fit an SPL to the track and cascade spectra in the energy range of 100 TeV to 1 PeV\footnote{While the energy ranges that IceCube uses for each of their track and cascade fits differ somewhat, the effect is marginal on the fits, but should be accounted for in any fit to the data.} and plot the apparent spectral index as a function of the coupling in fig.~\ref{fig:IC gamma}.
The feature at lower couplings comes from the atmospheric $g^{(\prime)}_{31}$ and $g^{(\prime)}_{32}$ couplings while the feature at higher couplings come from the solar $g^{(\prime)}_{21}$ couplings.
This is because the solar couplings lead to an effect at lower energies than the atmospheric couplings for a given value of $g^{(\prime)}_{ij}$ as seen in fig.~\ref{fig:invisible} and the top left panel of fig.~\ref{fig:visible}.
Then, since we are looking at a fixed energy range for IceCube, in order to keep $\Gamma L\sim1$ fixed, $g$ and $E$ must be changed together.
Thus for a fixed energy range the atmospheric features will occur at smaller couplings than the solar feature.

Fig.~\ref{fig:IC gamma} shows a somewhat stronger effect from invisible decay than visible decay.
This is because the depletion term is non-positive and the regeneration term is non-negative.
In the invisible case the flux is only lower than in the SM, while the visible case may partially cancel that out, depending on the parameters; compare e.g.~only $g^{(\prime)}_{32}>0$ in fig.~\ref{fig:invisible} and the top left panel of fig.~\ref{fig:visible}.
We note that this trend is generally true as the regeneration term can be considerably larger than the depletion term in some cases.

\begin{figure*}
\centering
\includegraphics[width=0.49\textwidth]{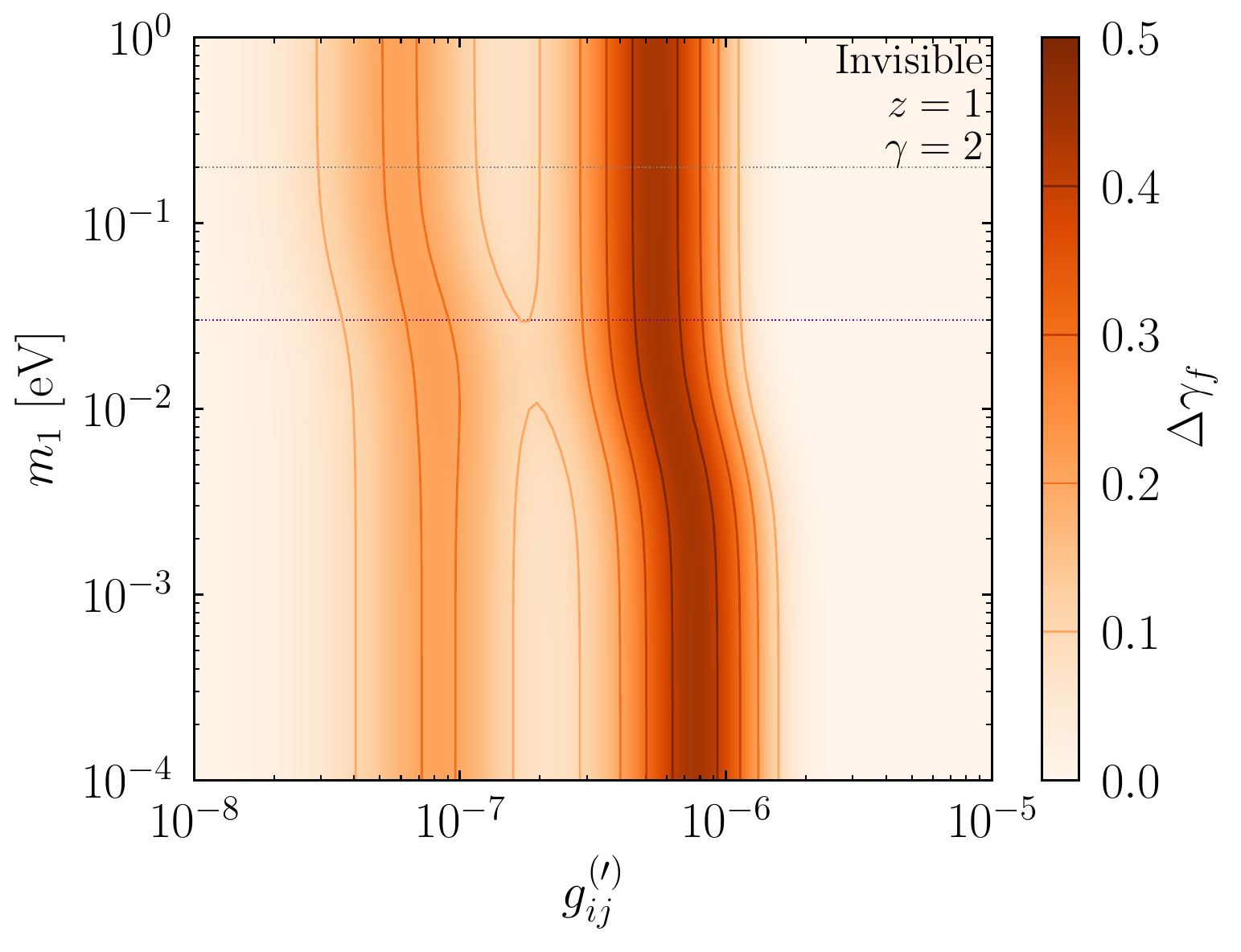}
\includegraphics[width=0.49\textwidth]{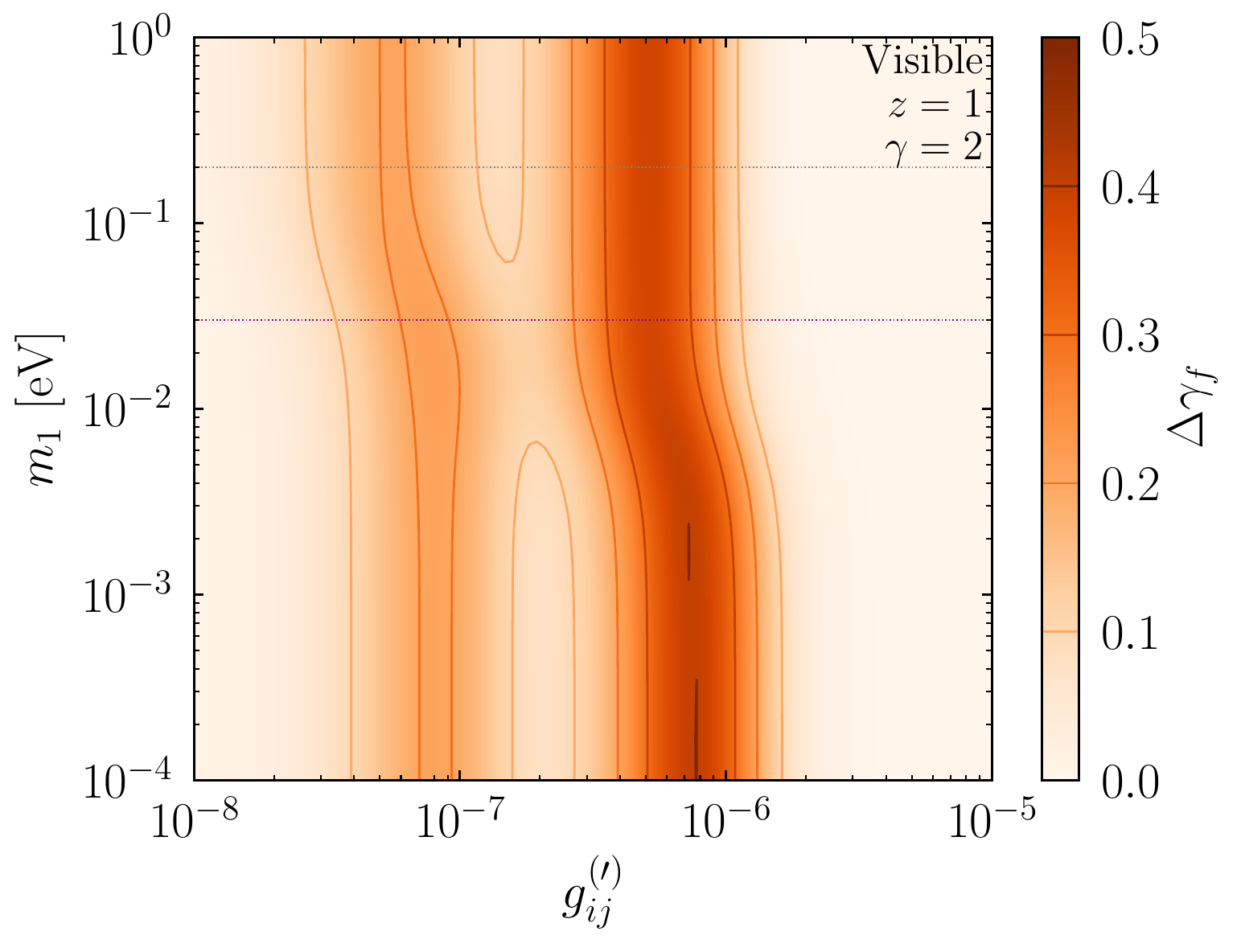}
\caption{The difference in cascade and track spectral indices at the Earth, $\Delta\gamma_f=\gamma_{f,c}-\gamma_{f,t}$, as a function of the lightest neutrino mass and the coupling assuming an initial flavor ratio $(1:2:0)$.
All six off-diagonal couplings are kept equal, the redshift distribution is $R(z)=\delta(z-1)$, and the initial spectral index is $\gamma=2$.
The purple line at $m_1=0.03$ eV is the 95\% CL upper limit from cosmology \cite{Aghanim:2018eyx}.
The current upper limit from KATRIN is just above the figure, but the gray line at $m_1=0.2$ eV is the projected 90\% CL upper limit \cite{Aker:2019uuj}.
\textbf{Left}: The invisible decay case with only the SM and depletion contributions which has $\max\Delta\gamma_f=0.44$.
\textbf{Right}: The visible decay case with the SM, depletion, and regeneration contributions which has $\max\Delta\gamma_f=0.40$.
Both figures continue down to $m_1=0$ as they are at $m_1=10^{-4}$ eV with no discernible change.}
\label{fig:IC gamma 2D}
\end{figure*}

To fully investigate this effect we plotted the difference in track and and cascade spectral indices, $\Delta\gamma_f\equiv\gamma_{f,c}-\gamma_{f,t}$ in fig.~\ref{fig:IC gamma 2D} for invisible and visible decay as a function of both the coupling and the mass scale.
We found that the initial spectral index has almost no effect on $\Delta\gamma_f$ for visible decay (it has exactly no effect for invisible decay) as it shifts the spectrum of both tracks and cascades nearly equally.
The maximum difference in final track and cascade spectral indices for various coupling configurations are listed in table \ref{tab:max delta gamma f}.
For invisible decay $\Delta\gamma_f\ge0$ always holds.
For visible decay it nearly only holds, however we find that for the cases when only $g^{(\prime)}_{21}$ ($g^{(\prime)}_{31}$) is non-zero that the minimum is $-0.011$ ($-0.01$).
That is, other than a few very small exceptions, it is \emph{always} the case that the cascade spectrum is softer than the track spectrum in the presence of neutrino decay.
This is conditioned upon the normal mass ordering, in the inverted ordering the situation is partially reversed.

\begin{table}
\centering
\caption{The maximum difference in the final track and cascade spectral indices as observed at the Earth where $\Delta\gamma_f=\gamma_{f,c}-\gamma_{f,t}$ for different couplings either one at a time or all three.}
\begin{tabular}{|c|c|c|c|c|}
\hline
$\max\Delta\gamma_f$&$g^{(\prime)}_{21}$&$g^{(\prime)}_{31}$&$g^{(\prime)}_{32}$&All\\\hline
Invisible&0.006&0.200&0.200&0.438\\
Visible&0.042&0.227&0.172&0.400\\\hline
\end{tabular}
\label{tab:max delta gamma f}
\end{table}

IceCube has reported a measurement of the track spectral index of $\gamma_{f,t}=2.13\pm0.13$ \cite{Aartsen:2016xlq} and a measurement of the cascade spectral index of $\gamma_{f,c}=2.67\pm0.07$ \cite{Niederhausen:2015svt} over somewhat different energy ranges.
A simple statistical test yields a difference of $\chi^2=11$ which can be interpreted as $3.3$ $\sigma$ tension and, in turn, $3.3$ $\sigma$ preference for invisible partial neutrino decay \cite{Denton:2018aml}.
To illustrate the preferred parameters the simple $\chi^2$ test statistic we use is,
\begin{equation}
\chi^2=\chi^2_{\rm track}+\chi^2_{\rm cascade}\,,
\end{equation}
where
\begin{align*}
\chi^2_{\rm track}&=\left(\frac{\gamma_{f,t}-2.13}{0.13}\right)^2\,,\\
\chi^2_{\rm cascade}&=\left(\frac{\gamma_{f,c}-2.67}{0.07}\right)^2\,.
\end{align*}
We take the track and cascade spectral indices over $E_f\in[100$ TeV$,1$ PeV$]$ as before which is slightly different from the experimental data but does not significantly affect our results.
We also considered a constraint on the sum of the neutrino masses from cosmology, 
$\chi^2_{m_1}=(\sum_im_i/0.06{\rm\ eV})^2$ based on the constraint from Planck TT, TE, EE+lowE+lensing+BAO data sets \cite{Aghanim:2018eyx} which is that $\sum_im_i<0.12$ eV at 95\% CL.
The effect on the preferred regions was negligible, so we fixed $m_1=0$.

\begin{figure*}
\centering
\includegraphics[width=0.49\textwidth]{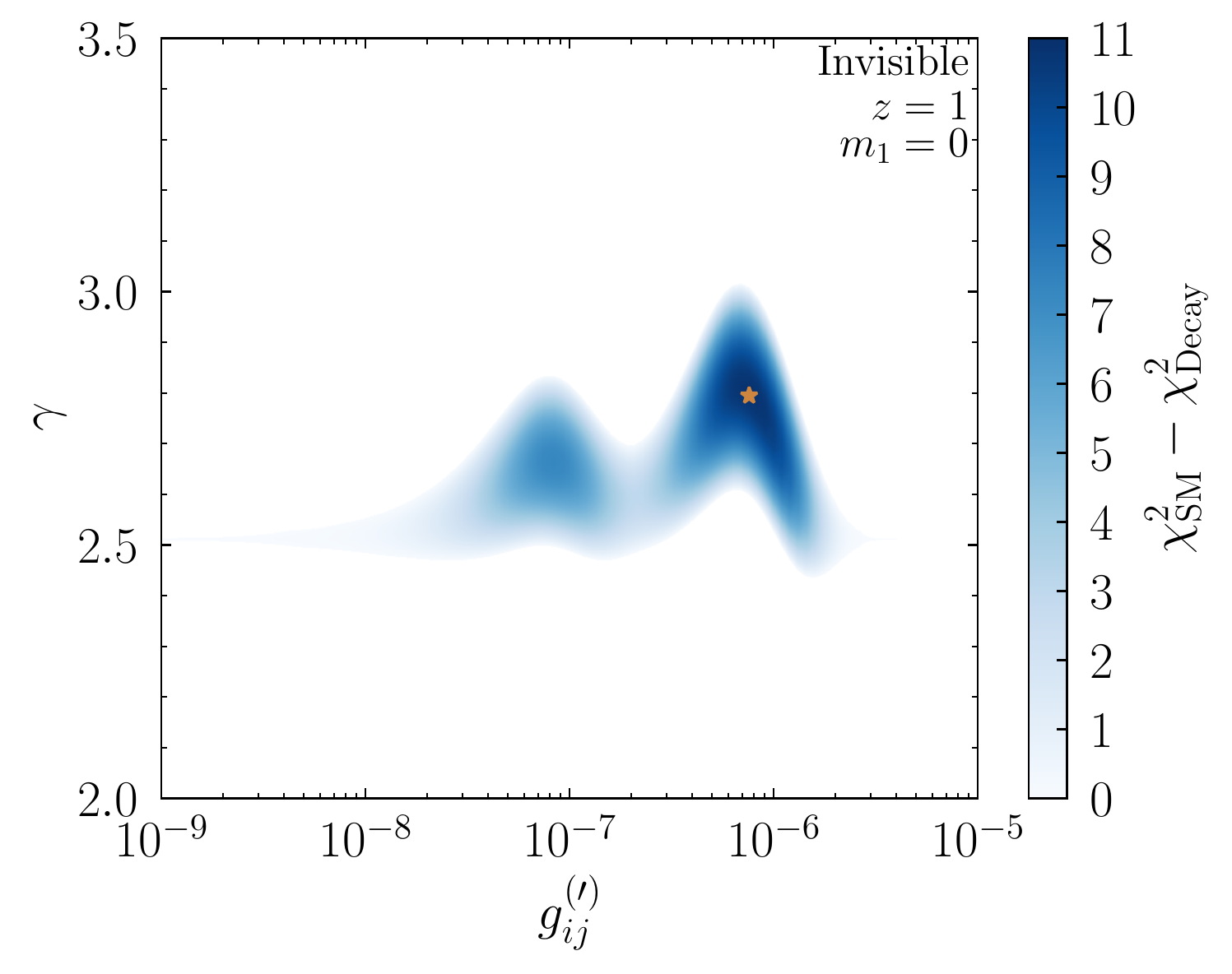}
\includegraphics[width=0.49\textwidth]{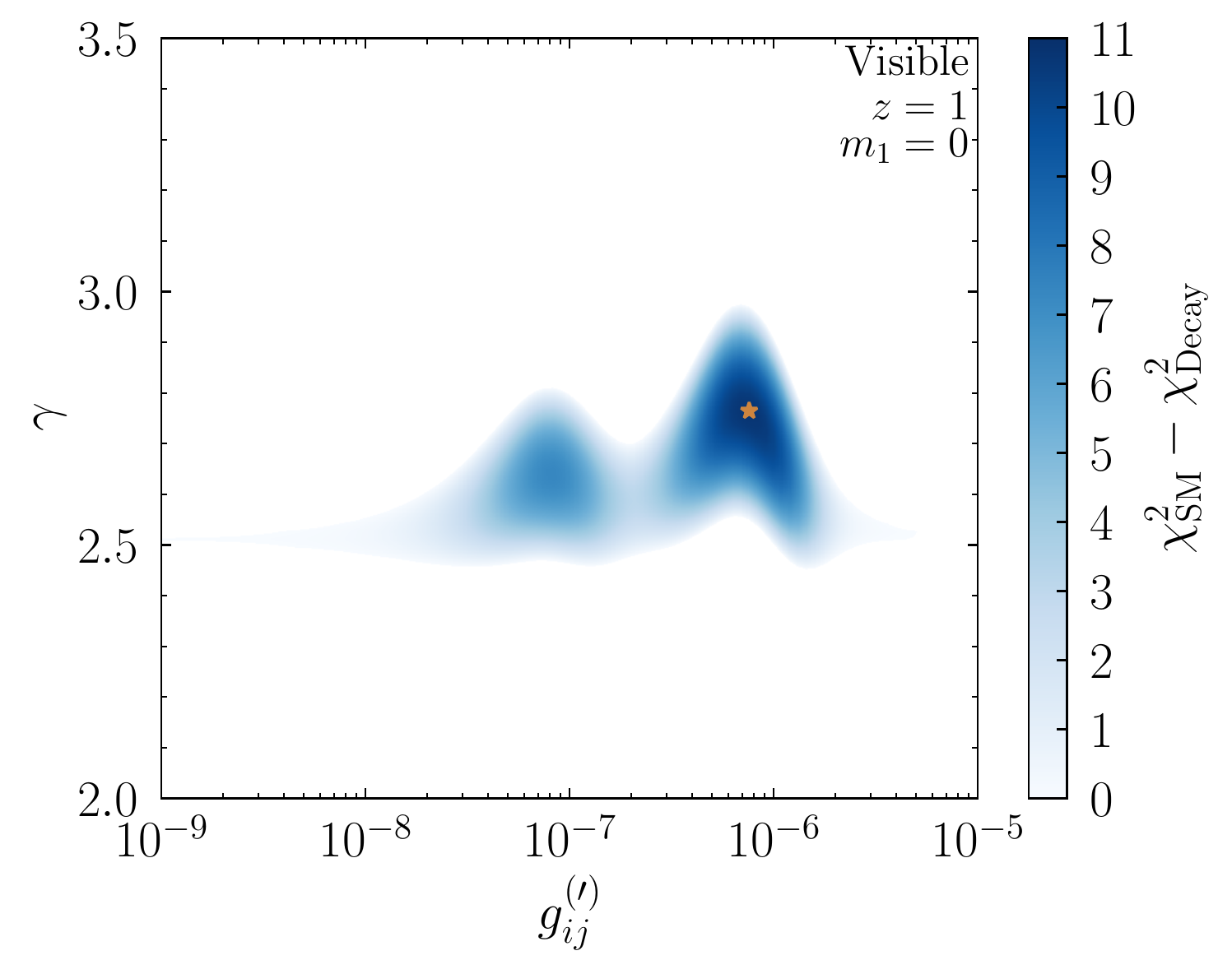}
\caption{The preferred regions of parameter space assuming both scalar and pseudo-scalar interactions for all channels, $R(z)=\delta(z-1)$, and assuming an initial flavor ratio $(1:2:0)$.
The best fit points, $(g_{ij}^{(\prime)},\gamma)$, denoted on the figures by the stars, for the invisible and visible decay cases are $(7.59\e{-7},2.80)$ and $(7.59\e{-7},2.77)$ respectively at which points the decay scenarios are preferred over the SM at $\Delta\chi^2=10.9$ and $10.8$ respectively.
White regions are disfavored relative to the SM.}
\label{fig:chisq}
\end{figure*}

The preferred regions for invisible and visible decay are shown in fig.~\ref{fig:chisq} for the benchmark decay parameters and varying the coupling (all six couplings are taken to be equal) and the initial spectral index.
Consistent with our previous simple estimate, we find some places where the $\chi^2$ is smaller in the case of decay than in the SM; these regions are indicated in blue with the darker blue regions being the regions that are most preferred.
In addition, we note that there is a horizontal line right on the edge of the allowed region for both large and small couplings in the fit at $\gamma=2.51$ which is the best fit spectral index in the SM.
The scenario of full decay might seem like an extreme one, but since the spectra of each flavor return to the same one as initially it is difficult to probe this.
While the flavor ratio is different in this scenario (see \ref{sec:analytic validate}), since the measured spectral indices are different any comparison with the flavor data only makes sense within a certain energy range.
For example, in the full decay scenario the flux at the Earth is pure $\nu_1$ which would predict a deficit of tracks and a dominantly $\nu_e$ flux.
This is compatible with the IceCube flavor data at lower energies which is essentially the reason why the data prefers a neutrino decay scenario over the SM.

Assuming that these spectra are confirmed with future data including e.g.~IceCube-Gen2, KM3NeT, and Baikal, we would find $\Delta\gamma_f=0.54$.
At the largest values of $\Delta\gamma_f$ from neutrino decay shown in table \ref{tab:max delta gamma f} of 0.44 (0.40) for invisible (visible) decay which leads to a consistent result at $\chi^2=0.14$ ($\chi^2=0.25$) which can be interpreted as $\sqrt{\Delta\chi^2}\to3.3$ $\sigma$ model preference in either case assuming Wilks' theorem.
A more detailed analysis including more spectral information than the slope for an SPL fit would be helpful in identifying the features of neutrino decay; these analyses require unfolding the final neutrino spectrum from the data under a given assumption and thus are difficult to perform given publicly available data.
In addition, a partial deficit in the $\nu_\tau$ flux at lower energies would be clear evidence of neutrino decay \cite{Denton:2018aml}.

\section{Discussion}
\label{sec:discussion}
While the effects of invisible neutrino decay are fairly straightforward -- a depletion of events below a certain energy -- and have been previously investigated, the effects of visible decay including a proper treatment of the spectra are much more complicated.
Visible decay depends not only on the couplings, neutrino energy, and the distance traveled, it also depends on the absolute neutrino mass scale, the initial neutrino spectrum, the nature of the couplings, and the nature of the masses of the neutrinos.
We now discuss the several trends seen in the previous sections.

We see that the higher the mass scale, the larger the flux.
This is because when $m_1$ is larger, the final neutrinos have energies closer to the initial neutrinos.
Since we are considering steeply falling spectra, this results in a larger effect from regeneration compared with a smaller mass scale wherein the final neutrinos may have much lower energy and become swamped out by the steeply falling flux.
In the limit $m_1\to\infty$ we recover the case where $E_f=E_i$ even for regeneration since $x_{ij}\to1$ in this limit and the range of the integral over initial (and intermediate for the case with two decays) energy becomes vanishingly small.
However, as shown in fig.~\ref{fig:visible}, even for the largest physically allowed values of $m_1$ from cosmology deviations from this limit already exist.

The source spectral index also has a very large effect on the flux.
If the spectrum is very hard (say, $\gamma=1$ within IceCube's region of interest) then the regeneration component is extremely large.
Softer and softer spectra result in less contribution from the regeneration term and, as $\gamma\to\infty$ we would recover the invisible decay case, although this is not a particularly physical limit.
If, however, the neutrino mass scale is arbitrarily large, ($m_1\gg1$ eV) then the regeneration term is independent of the initial spectrum.
These features are shown analytically in section \ref{sec:analytic} without cosmology, but the trends still hold with cosmology.

We also investigated the effect of scalar interactions, pseudo-scalar interactions, or both and found only small differences among those cases\footnote{The reason both scalar and pseudo-scalar interactions appears to be different is because the width is now approximately double the case of either one.}.
In addition, the channel $\nu\to\nu$ is quite similar to the $\nu\to\bar\nu$ channel.
The first channel is relevant for Dirac neutrinos as it is the only one that contributes and the second is relevant for Majorana neutrinos in the case where one can differentiate neutrinos and antineutrinos in the source.
Since IceCube has quite limited capabilities to differentiate neutrinos from antineutrinos, we consider both simultaneously: $\nu\to\nu$ and $\nu\to\bar\nu$.
This also has the added benefit of ensuring that the double decay case $\nu_3\to\nu_2\to\nu_1$ is handled correctly.

To summarize, we found that the initial spectrum $\gamma$, the neutrino mass scale given by $m_1$, and the texture of the matrices $g_{ij}^{(\prime)}$ can have a significant effect, while effects related to the nature of the coupling or the redshift evolution have only marginal effect.

Depending on the couplings the effect of neutrino decay falls into one of three categories.
For small couplings such that $\Gamma L\ll1$ in the energy and distance ranges of interest, the SM is recovered.
For large couplings such that $\Gamma L\gg1$ in the energy and distance ranges of interest, we enter the full decay scenario where the spectrum is the same as in the SM, but the relative normalizations of the different flavors will be different depending on which $g^{(\prime)}_{ij}$ couplings are large.
For cases when the $\Gamma L\sim1$ for the relevant energies and distances, we have partial neutrino decay which, in addition to modifying the normalizations of each flavor, can also have significant effects on the shape for each flavor.
This opens up the possibility for detection under the reasonable assumption that astrophysics and standard oscillations cannot reproduce such an effect.
We see in figs.~\ref{fig:IC gamma 2D}-\ref{fig:chisq} that the expected allowed regions in parameter space for invisible and visible decays are fairly similar.
It is, in principle, possible to differentiate between the two, but only under certain assumptions.
If the spectra of two (or three) different flavors is measured in detail and they are found to recover the same spectrum at high energies, visible and invisible decay can be distinguished if one assumes that the initial source spectrum continues down throughout the entire detection regime since invisible decay requires a softer initial spectrum than visible decay to fit the same data.
Alternatively, a very fine grained measurement could reveal a slight bump like structure in the spectrum (see many of the curves in fig.~\ref{fig:visible}, although this could be misinterpreted as evidence for dark matter decaying or annihilating to neutrinos \cite{Ahlers:2015moa,Chianese:2017lvv,Chianese:2017nwe,Chianese:2017lxm,Denton:2018aml,Aartsen:2018mxl,Chianese:2018ijk,Sui:2018bbh,Chianese:2019kyl,Chianese:2019xro,Chianese:2019zkn,Dekker:2019gpe}).

The effect of neutrino decay can be identified at IceCube and is a possible explanation \cite{Denton:2018aml} for the $\gtrsim3$ $\sigma$ tension in the track to cascade spectra \cite{Aartsen:2016xlq,Denton:2018aml,Aartsen:2018vez,Palladino:2019pid}.
If the spectra continue to disagree with more data, this may become compelling evidence for neutrino decay.
A careful investigation of the spectrum of each flavor, including new information about $\nu_\tau$'s, would be necessary to determine the particular parameters, in particular looking at the texture of the $g_{ij}$ and $g'_{ij}$ matrices.
It is very interesting to note that neutrino decay in the normal mass ordering predicts a cascade spectrum that is softer than that of tracks; if IceCube had measured the reverse then neutrino decay would not explain the data (or we would have to invoke neutrino decay in the inverted mass ordering).
If such a signal of neutrino decay is confirmed, we anticipate it to have couplings of $g^{(\prime)}_{3j}\sim10^{-7}$ and $g^{(\prime)}_{21}\sim10^{-6}$ as shown in fig.~\ref{fig:chisq} in order to have the desired feature within IceCube's region of interest subject to the details of the true redshift evolution of the flux.
We also see in fig.~\ref{fig:chisq} that the preferred initial spectral indices are softer in the presence of neutrino decay (2.80 and 2.77 for invisible and visible decay respectively) than in the SM (2.51).
This is consistent with expectations as neutrino decay tends to harden a spectrum so matching the data requires a softer initial spectrum when decay is included.
Moreover, we see that invisible decay prefers a slightly softer initial spectrum than visible which is due to the regeneration term in visible decay partially canceling out the hardening from the depletion term.

While next-generation experiments such as GRAND and POEMMA \cite{Alvarez-Muniz:2018bhp,Olinto:2017xbi} (in addition to current experiments such as ANITA \cite{Gorham:2008dv}) will have unique sensitivity to tau neutrinos thus providing important flavor information that is generally difficult for any experiment (including IceCube) to detect, their higher energy thresholds makes them sensitive only to larger couplings than IceCube.
Due to the facts that they will only be measuring a single flavor and the source flux is relatively unknown, disentangling neutrino decay features from astrophysical features will be extremely difficult.

\section{Conclusions}
\label{sec:conclusions}
Visible neutrino decay has a rich phenomenology that can be probed at IceCube by simultaneously measuring the flavor and energy of the high energy astrophysical neutrino flux.
Due to the extremely long distances the neutrinos at IceCube travel en route to the Earth, IceCube provides one of the most sensitive direct probes of neutrino decay itself.
In addition, while the initial flux is unknown, IceCube compensates for this by detecting all three flavors and partially differentiating among them.
In this article we have calculated the flavor transition probabilities for all channels as a function of the initial neutrino spectrum, the absolute neutrino mass scale, and the various couplings for the first time.
This calculation involves an integral over the distance at which the decay happens and the initial neutrino energy which is larger than the final neutrino energy.
We have also included the effect of multiple decays involving extra integrals over the second decay point and the energy of the intermediate neutrino.
Finally, we incorporated the effect of cosmology and the expansion of the universe.

Neutrino decay in the partial decay regime modifies the spectrum of neutrinos seen at IceCube differently for different flavors, while in the SM there can be nearly no difference.
This makes IceCube a powerful probe for neutrino decay.
Since IceCube has measured this flux and has some sensitivity to flavors, it provides a powerful probe of neutrino decay.
As IceCube sees some hints at $\sim3.3$ $\sigma$ that the spectra may be different for different flavors, neutrino decay offers a mechanism for explaining this difference.
The fact that it is the cascade spectrum that is softer than the track spectrum is consistent with neutrino decay in the normal mass ordering.
The preferred neutrino decay parameters are shown in fig.~\ref{fig:chisq} and more detailed measurements from IceCube, KM3NeT, and Baikal can further test neutrino decay in coming years.

\begin{acknowledgments}
We thank Mauricio Bustamante, Julia Gehrlein, Matheus Hostert, Kristian Moffat, Stephen Parke, and Anna Suliga for helpful comments.
AA and PBD thank the Fermilab theory department for their hospitality.
AA is funded by the UKRI STFC, and has received funding/support from the European Union’s Horizon 2020 research and innovation program under the Marie Skłodowska - Curie grant agreements No. 690575 (RISE InvisiblesPlus) and No. 674896 (ITN Elusives). 
PBD is supported by the US Department of Energy under Grant Contract DE-SC0012704.
PBD acknowledges support from the Fermilab Neutrino Physics Center.
\end{acknowledgments}

\onecolumngrid

\appendix
\section{Non-Oscillation Averaged Transition Probabilities}\label{App. sec. non-avg}
When dealing with decays over local distances, such as is relevant for long-baseline, atmospheric, or reactor neutrinos, it is necessary to account for the interference of the decay amplitudes, as the neutrinos involved will not have decohered before decaying.
In sec.~\ref{ssec:invisible} and \ref{ssec:visible decay}, we gave the expressions for the oscillation averaged probabilities.
Below we compute the full expressions for the depletion and regeneration components in the flavor basis and give the full non-oscillation averaged results.

\subsection{Depletion Component}\label{App. ssec. dep}
From eq.~\ref{eq.inv}, we have
\begin{align}
P_{\alpha\beta}^{\text{inv}}\left(E,L\right)&= \bigg|U^*_{\alpha1}U_{\beta_1} + U^*_{\alpha2}U_{\beta_2}e^{-i\frac{\Delta m^2_{21}}{2E}L}e^{-\frac{1}{2}\Gamma_2 L} + U^*_{\alpha3}U_{\beta_3}e^{-i\frac{\Delta m^2_{31}}{2E}L}e^{-\frac{1}{2}\Gamma_3 L}\bigg|^2\, \nonumber \\
&= P_{\alpha\beta}^{\text{SM}} -|U_{\alpha 2}|^2 |U_{\beta 2}|^2 \left(1-e^{-\Gamma_2 L}\right) - |U_{\alpha 3}|^2 |U_{\beta 3}|^2 \left(1-e^{-\Gamma_3 L}\right) \, \nonumber \\
&- 2\sum_{i>j}\left|U^*_{\alpha i}U_{\beta i}U_{\alpha j}U^*_{\beta j}\right|\cos\left(\phi_{ij}+\frac{\Delta m^2_{ij}L}{2E}\right)\left(1-e^{-\frac{1}{2}(\Gamma_i + \Gamma_j)L}\right)\,,
\end{align}  
where $\phi_{ij}=\arg\left(U^*_{\alpha i}U_{\beta i}U_{\alpha j}U^*_{\beta j}\right)$ are CP violating phases.
\subsection{Regeneration Component}\label{App. ssec. reg}
Starting with eq.~\ref{eq:Amp}, we reinsert the phases and move to the flavor basis. We can construct the full regeneration transition probability as,
\begin{align}
P_{\alpha\beta}^{\rm reg}(E_i,E_f,L) &= \int_{0}^LdL_1 \Delta P_{\alpha\beta}^{\rm reg}(E_i,E_f,L,L_1) + \int_0^LdL_1\int_{L_1}^LdL_2 \Delta P_{\alpha\beta}^{\rm reg,2}(E_i,E_f,L,L_1,L_2)\, \nonumber \\
&= \int_{0}^LdL_1 |\sum_{i>j} U^*_{\alpha i} U_{\beta j}\mathcal A_{ij}^{\rm reg}|^2\,  + \int_0^LdL_1\int_{L_1}^LdL_2 |U_{\alpha 3}|^2|U_{\beta 1}|^2|A_{31}^{\rm reg,2}|^2.
\end{align}
As the decay $\nu_3\to\nu_2\to\nu_1$ does not interfere with the single decays, it does not contribute to the oscillations. We can therefore compute the first term, and simply add on the expression given in eq.~\ref{eq:2decays}.
\begin{align}
    &\int_{0}^L dL_1 |\sum_{i>j} U^*_{\alpha i} U_{\beta j}\mathcal A_{ij}^{\rm reg}|^2 = 
     |U_{\alpha3}|^2 |U_{\beta1}|^2 \Gamma_{31}W_{31} \int^L_0 e^{-\Gamma_3 L_1} dL_1
    + |U_{\alpha3}|^2 |U_{\beta2}|^2 \Gamma_{32}W_{32}e^{-\Gamma_2 L}\int^L_0 e^{-(\Gamma_3 - \Gamma_2) L_1} dL_1\, \nonumber \\
    &+ |U_{\alpha2}|^2 |U_{\beta1}|^2 \Gamma_{2}W_{21} \int^L_0 e^{-\Gamma_2 L_1} dL_1 
    + 2\mathfrak{R}\bigg\{|U_{\alpha3}|^2 U_{\beta1}U^*_{\beta2}\sqrt{\Gamma^*_{32}W^*_{32}\Gamma_{31}W_{31}}e^{-\frac{1}{2}\Gamma_2L}e^{i\frac{\Delta m^2_{21}}{2E_f}L}\int^L_0 e^{-\big(i\frac{\Delta m^2_{21} }{2E_f} + \frac{1}{2}(2\Gamma_3 -\Gamma_2)\big)L_1} dL_1 \, \nonumber\\
    &+ |U_{\beta1}|^2 U^*_{\alpha3}U_{\alpha2}\sqrt{\Gamma^*_{21}W^*_{21}\Gamma_{31}W_{31}}\int^L_0 e^{-\big(i\frac{\Delta m^2_{32} }{2E_i} + \frac{1}{2}(\Gamma_3 +\Gamma_2)\big)L_1} dL_1\, \nonumber \\
    &+ U^*_{\alpha3}U_{\alpha2}U_{\beta2}U^*_{\beta1}\sqrt{\Gamma^*_{32}W^*_{32}\Gamma_{21}W_{21}}e^{-\frac{1}{2}\Gamma_2L}e^{-i\frac{\Delta m^2_{21}}{2E_f}L}\int^L_0 e^{-\big(i\big(\frac{\Delta m^2_{32}}{2E_i} - \frac{\Delta m^2_{21}}{2E_f}\big) +\frac{1}{2}\Gamma_3\big)L_1}dL_1\bigg\}.
\end{align}
To evaluate the integrals, we use that,
\begin{equation}
    \int^L_0 e^{-(ai+b)L_1}dL_1 = \frac{1}{ai+b}\big(1 - e^{-(ai+b)L}\big)\,,
\end{equation}
and to obtain the real parts,
\begin{multline}
\mathfrak{R}\left[\frac{ze^{-(ci+d)L}}{ai+b}\left(1 - e^{-(ai+b)L}\right)\right] = \frac{|z|}{a^2 + b^2}\bigg[e^{-dL}\left(b\cos\left(\phi - cL\right)+ a\sin\left(\phi - cL\right)\right)\, \\
-e^{-\left(b+d\right)L}\left(b\cos\left(\phi - \left(a+c\right)L\right)+ a\sin\left(\phi -\left(a+c\right)L\right)\right)\bigg]\,,
\end{multline}
where $\phi=\arg\left(z\right)$.
After some algebra, we obtain
\begin{align}
&P_{\alpha\beta}^{\rm reg}(E_i,E_f,L)= |U_{\alpha3}|^2 |U_{\beta1}|^2 \bigg[\frac{\Gamma_{31}W_{31}}{\Gamma_3}\big(1-e^{-\Gamma_3 L}\big) + \frac{\Gamma_{32}W_{32}\Gamma_{21}W_{21}}{\Gamma_3-\Gamma_2}\left(\frac{1}{\Gamma_2}\left(1-e^{-\Gamma_2 L}\right) - \frac{1}{\Gamma_3}\left(1-e^{-\Gamma_3 L}\right)\right)\bigg]\, \nonumber\\ 
&+ |U_{\alpha3}|^2 |U_{\beta2}|^2\bigg[\frac{\Gamma_{32}W_{32}}{\Gamma_3 - \Gamma_2} \big(1-e^{-(\Gamma_3-\Gamma_2) L}\big)\bigg]
+ |U_{\alpha2}|^2 |U_{\beta1}|^2\bigg[\frac{\Gamma_{21}W_{21}}{\Gamma_2} \big(1-e^{-\Gamma_2 L}\big)\bigg] \, \nonumber\\
&+A\bigg[e^{-\frac{1}{2}\Gamma_2 L}\bigg(\frac{1}{2}(2\Gamma_3 - \Gamma_2)\cos\bigg(\phi + \frac{\Delta m^2_{21}}{2E_f}L\bigg) + \frac{\Delta m^2_{21}}{2 E_f}\sin\bigg(\phi + \frac{\Delta m^2_{21}}{2E_f}L\bigg)\bigg)- e^{-\Gamma_3 L}\bigg(\frac{1}{2}(2\Gamma_3 - \Gamma_2)\cos{\phi} + \frac{\Delta m^2_{21}}{2E_f}\sin{\phi}\bigg)\bigg]\, \nonumber\\
&+ B\bigg[\bigg(\frac{1}{2}(\Gamma_3 + \Gamma_2)\cos{\xi} + \frac{\Delta m^2_{32}}{2E_i}\sin{\xi}\bigg)- e^{-\frac{1}{2}(\Gamma_3+\Gamma_2) L}\bigg(\frac{1}{2}(\Gamma_3 + \Gamma_2)\cos\bigg(\xi - \frac{\Delta m^2_{32}}{2E_i}L\bigg) + \frac{\Delta m^2_{32}}{2E_i}\sin\bigg(\xi - \frac{\Delta m^2_{32}}{2E_i}L\bigg)\bigg)\bigg]\, \nonumber\\
&+ C\bigg[e^{-\frac{1}{2}\Gamma_2 L}\bigg(\frac{1}{2}\Gamma_3 \cos\bigg(\psi - \frac{\Delta m^2_{21}}{2E_f}L\bigg) + \bigg(\frac{\Delta m^2_{32}}{2E_i} - \frac{\Delta m^2_{21}}{2E_f}\bigg)\sin\bigg(\psi - \frac{\Delta m^2_{21}}{2E_f}L\bigg)\bigg)\, \nonumber\\
&-e^{-\frac{1}{2}(\Gamma_3 + \Gamma_2)L}\bigg(\frac{1}{2}\Gamma_3 \cos\bigg(\psi - \frac{\Delta m^2_{32}}{2E_i}L\bigg) + \bigg(\frac{\Delta m^2_{32}}{2E_i} - \frac{\Delta m^2_{21}}{2E_f}\bigg)\sin\bigg(\psi - \frac{\Delta m^2_{32}}{2E_i}L\bigg)\bigg)\bigg]\,,
\end{align}
where $A, B, C$ are prefactors, and $\phi$, $\xi$, $\psi$ are CP violating phases, given as:
\begin{align*}
    A &=\frac{8E^2_{f}|U_{\alpha3}|^2|U_{\beta1}U^*_{\beta2}\sqrt{\Gamma^*_{32}W^*_{32}\Gamma_{31}W_{31}}|}{E^2_f (2\Gamma_3 - \Gamma_2)^2 +(\Delta m^2_{21})^2 }\,, & \phi&= \arg(U_{\beta1}U^*_{\beta2}\sqrt{\Gamma^*_{32}W^*_{32}\Gamma_{31}W_{31}}) \\
    B &=\frac{8E^2_{i}|U_{\beta1}|^2 |U_{\alpha2}U^*_{\alpha3}\sqrt{\Gamma^*_{21}W^*_{21}\Gamma_{31}W_{31}}|}{E^2_{i}(\Gamma_3 + \Gamma_2)^2 +(\Delta m^2_{32})^2}\,, & \xi&= \arg(U_{\alpha2}U^*_{\alpha3}\sqrt{\Gamma^*_{21}W^*_{21}\Gamma_{31}W_{31}}) \\
    C &=\frac{8E^2_{i}E^2_{f}|U^*_{\alpha3} U_{\alpha2}U_{\beta2}U^*_{\beta1}\sqrt{\Gamma^*_{21}W^*_{21}\Gamma_{32}W_{32}}|}{E^2_{i}E^2_{f}\Gamma^2_3 +(E_{f}\Delta m^2_{32} - E_{i}\Delta m^2_{21})^2}, & \psi&= \arg(U^*_{\alpha3} U_{\alpha2}U_{\beta2}U^*_{\beta1}\sqrt{\Gamma^*_{21}W^*_{21}\Gamma_{32}W_{32}})\,.
\end{align*}
Note that we have not integrated over the initial, or intermediate, energies here. 
We would have to integrate the terms separately, as described in sec.~\ref{ssec:visible decay}, as the integration limits would vary according to the decay in question. 

\section{Validation of the Analytic Expression}
\label{sec:analytic validate}
\begin{figure}
\centering
\includegraphics[width=0.5\textwidth]{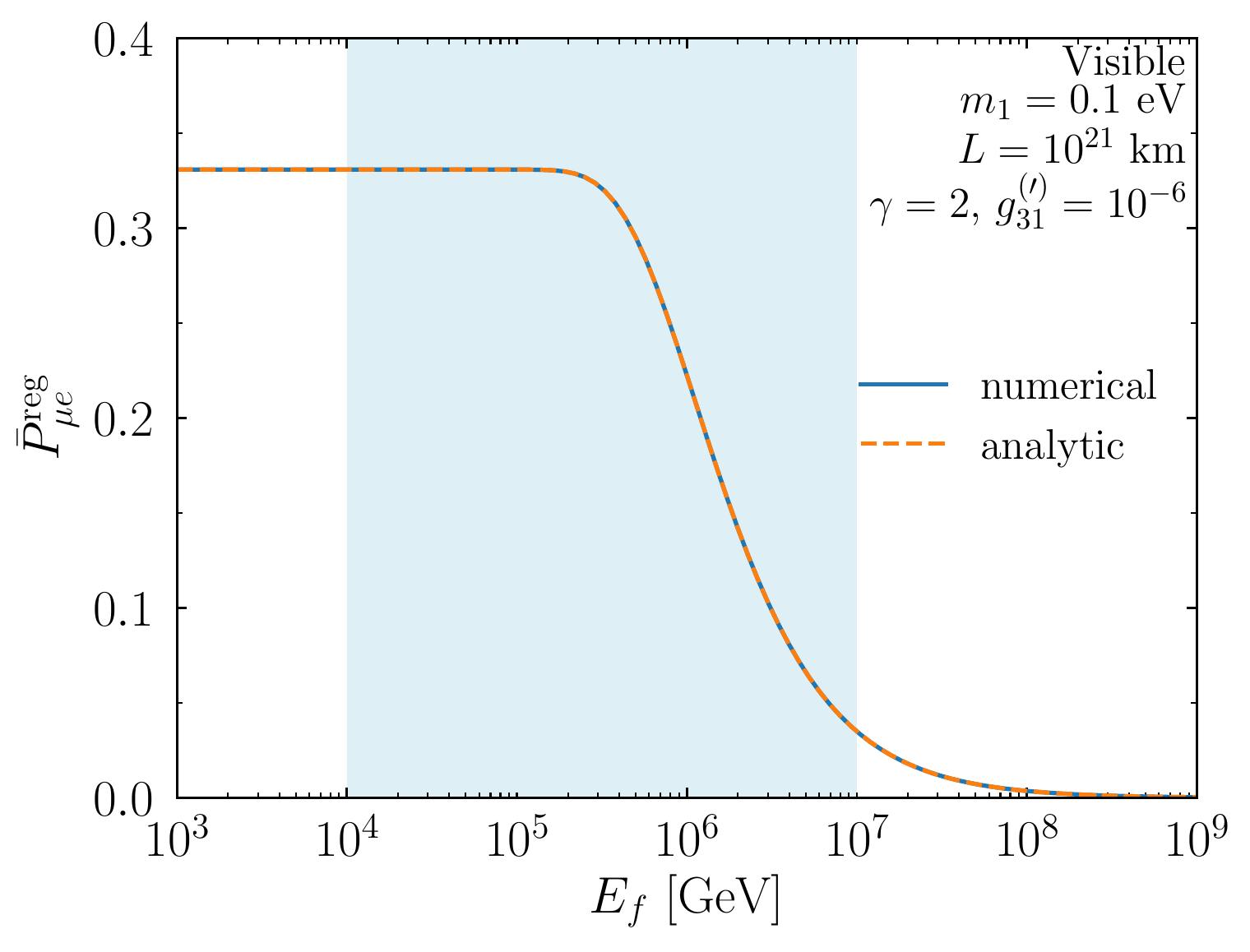}
\caption{The regeneration component of the $\nu_\mu\to\nu_e$ probability calculated for the benchmark parameters except without cosmology and with only $g_{31}$ and $g'_{31}$ non-zero.}
\label{fig:analytic validate}
\end{figure}

Throughout this work all of the results are calculated by numerically integrating.
In fig.~\ref{fig:analytic validate} we confirm that the analytic expression in eq.~\ref{eq:analytic} agrees with the numerical result.

\twocolumngrid

\bibliography{Visible}

\end{document}